\documentclass[useAMS,usenatbib]{mn2e}
\pdfoutput=1

\usepackage{graphicx}
\usepackage{multirow}
\usepackage[centertags]{amsmath}
\usepackage{color}
\usepackage{amsfonts}
\usepackage{amssymb}

\title[Modes identification in rotating SPB stars]{Identification of pulsational modes in rotating slowly pulsating B-type stars}

\author[W. Szewczuk and J. Daszy\'nska-Daszkiewicz]{W. Szewczuk$^{1}$\thanks{E-mail:
szewczuk@astro.uni.wroc.pl (WS);\newline daszynska@astro.uni.wroc.pl (JDD)} and
J. Daszy\'nska-Daszkiewicz$^{1}$\footnotemark[1]{}\\
$^{1}$Instytut Astronomiczny, Uniwersytet Wroc{\l}awski,
    Kopernika 11, 51-622 Wroc{\l}aw, Poland}

\begin{document}

\date{}

\pagerange{\pageref{firstpage}--\pageref{lastpage}} \pubyear{}

\maketitle

\label{firstpage}

\begin{abstract}
Knowledge of the geometry of pulsational modes is a prerequisite for seismic modelling of stars.
In the case of slowly pulsating B-type (SPB) pulsators, the simple zero-rotation approach so far used for mode identification is usually not valid
because pulsational frequencies are often of the order of the rotational frequency.
Moreover, this approach allows us to determine only the spherical harmonic degree, $\ell$, while the azimuthal order, $m$,
is beyond its reach. On the other hand, because of the density of oscillation spectra of SPB stars, knowledge of $m$
is indispensable if one wants to assign the radial order, $n$, to the observed frequency peaks.
Including the effects of rotation via the traditional approximation, we perform identification
of the mode angular numbers ($\ell,~m$) for 31 SPB stars with available multicolour time series photometry.
Simultaneously, constraints on the rotational velocity, $V_{\rm rot}$, and the inclination angle, $i$,
are determined assuming uniform rotation and a constant value of $V_{\rm rot}\sin i$.
Dependence of the results on the adopted model is tested using HD\,21071 as an example.
Despite some model uncertainties and limitations of the method, our studies show
the correct approach to identifying the low-frequency oscillation modes.
\end{abstract}


\section{Introduction}

Slowly pulsating B-type (SPB) stars are main-sequence objects pulsating in high-order gravity modes.
The observational properties of the SPB stars were broadly discussed, e.g., by
\citet{Balona2000} and \citet{DeCat2002ASPC_259_A,DeCat2007CoAst.150..167D}.
These modes are driven by the classical $\kappa$-mechanism operating in the metal opacity bump
\citep*{DMP1993, Gautschy1993}.
There are three basic problems which complicate the interpretation of the SPB pulsations and
make seismic modelling of these pulsators particularly difficult.
First, their oscillation spectra obtained from ground-based observations are usually sparse and lack equidistant patterns.
Therefore, the only way to identify pulsation modes, i.\,e., to determine the spherical harmonic degree, $\ell$,
and the azimuthal order, $m$, is the use of the information contained in the light and/or line profile variations.
The exceptions are the stars observed from space, which have oscillation spectra with hundreds of frequency peaks
and with some clear structures \citep*[e.\,g.][]{Degroote2010,Balona2011,McNamara2012,Papics2012,Papics2014}.

The most popular tools derived from ground-based observations and
used for mode identification are the amplitude ratios and phase differences in various photometric passbands
\citep{Balona1979, Stamford1981, Watson1988}, because these data are much more easily obtained than the line profiles.
The amplitude ratios and phase differences in different passbands were called the photometric non-adiabatic observables by \citet*{Cugier1994},
who improved this method for $\beta$ Cephei stars by including non-adiabatic effects in their pulsational computations.
If all effects of rotation are ignored, these observables are independent of the azimuthal order, $m$, and inclination angle, $i$,
and modes with different harmonic degrees, $\ell$, occupy disjoint regions on the amplitude ratio versus phase difference diagrams.
The results for non-rotating SPB stellar models were presented for the first time by \citet{Townsend2002}.

It should be mentioned that the {mode degree, $\ell$, and}
azimuthal order, $m$, can be extracted from spectroscopic observations {as well}.
For details on the spectroscopic methods, we refer to, e.\,g.\,\citet*{Campos1980a,Campos1980b,
Balona1986a,Balona1986b,Aerts1992,CJDD2001,Gies1988,Kennelly1996,
Telting1997,Schrijvers1999,Schrijvers2002} and \citet*{Zima2006}.
Some of these methods take into account the effects of the Coriolis force
\citep[e.g.,][]{Telting1997,Schrijvers1999,Schrijvers2002, Zima2006}
but only to the first order and therefore they do not apply to low-frequency g modes.
The traditional approximation was applied only in the moment method by \citet{DeCat2005}
to a few SPB stars. The most recent overview of the spectroscopic methods can be found in \citet{Uytterhoeven2014}.

The second problem is associated with a very high density of the theoretical oscillation spectra of SPB stellar models.
For this reason, the assignment of the radial order, $n$, to the individual observed peaks is usually impossible
or can lead to the misinterpretation of the data. For example, \citet{Degroote2010, Degroote2012}  claimed
that they have discovered a series of dipole zonal modes equidistant in period in the SPB star HD\,50230.
Our findings, however, contradict this result \citep*{WSJDDWD2014}.
We showed that this equidistant pattern is rather accidental and its interpretation in terms of the asymptotic theory is dubious.

The third problem with the interpretation of pulsations in SPB stars is rotation,
which significantly changes, both, the equilibrium model and pulsational properties.
In the case of these pulsators, rotation plays an especially important role because even
at slow rotation pulsational frequencies can be of the order of the rotational frequency
and the effect of the Coriolis force cannot be neglected.
On the other hand, the effect of centrifugal deformation on g modes is small \citep{Ballot2012}.
The influence of rotation on long-period g modes was examined relying on the so-called traditional approximation
\citep*[e.\,g.][]{Lee1997, Townsend2003a, Townsend2003b, Townsend2005, JDD2007}
or on the truncated expansion in Legendre functions for the eigenfunctions \citep[e.\,g.][]{Lee1989, Lee2001}.

So far, the mode identification for SPB stars has been made mostly in the zero-rotation approximation,
e.g., by \citet{Townsend2002} and \citet{DeCat2004ASPC_310}, and the results indicated that most of these stars
pulsate in dipole modes.
The only SPB star for which the effects of rotation were included in the photometric method of mode identification was $\mu$ Eri
\citep*{JDD2008, JDD2015}. This identification was based on linear non-adiabatic calculations
with the effects of rotation treated in the traditional approximation.

Therefore, there is a need for this type of studies and the main goal of the present work is to make mode identification
with the effects of rotation taken into account for all SPB stars with available and appropriate photometric and/or spectroscopic data.
Uniform rotation is assumed.
We follow the method used in \citet{JDD2008, JDD2015}, which, in addition, allows us to constrain also the rotational velocity of our programme stars.
For comparison, we make also mode identification in the zero-rotation approximation.

The structure of the paper is as follows.
In Section\,2, we present observational data of selected SPB stars gathered from the literature,
and the luminosities and evolutionary masses determined by us.
In Section\,3, the methods of mode identification we use,
dependence on model parameters and the results are discussed.
We end with a summary in the last section.

\section{Programme stars}

We selected 31 SPB stars for which the data needed to put a star on the H--R diagram and perform mode identification were available.
The HD numbers of the stars and their stellar parameters are given in Table\,\ref{SPB_param}.
All the parameters, except the masses and luminosities, were taken from the literature.
If there were no determinations of the metallicity or colour excess, the solar value $Z_\odot=0.0134$ \citep{Asplund2009}
and 0\,mag were assumed, respectively.
In the case of HD\,28114, the colour excess, $E(B-V)$, in the Johnson filters
was computed from the colour excess, $E(b-y)$, in the Str\"omgren filters
\citep{Vogt1998}. For the star HD\,179588, \citet{Grillo1992} give the value of the interstellar extinction
in the Johnson $V$ filter, $A_V$. To obtain $E(B-V)$, we assumed the standard value of the ratio of
total to selective extinction, equal to 3.1.

The values of luminosity were determined in this paper using the bolometric corrections
from \citet{Flower1996}. Stellar masses were derived from evolutionary tracks
and they usually correspond to the values of luminosity and effective temperature from the centre of the error box.
There are three exceptions: HD\,74195, HD\,85953 and HD\,123515.
The  masses of these stars come from models which lie on the left bottom edges of the error boxes,
so that they have the following values of the effective temperature and luminosity:
$\log T_\mathrm{eff}$=4.243, $\log L/\mathrm L_\odot$=3.323 for HD\,74195,
$\log T_\mathrm{eff}$=4.312, $\log L/\mathrm L_\odot$=3.833 for HD\,85953, and
$\log T_\mathrm{eff}$=4.081, $\log L/\mathrm L_\odot$=2.484 for HD\,123515.
This is because the models from the centres of the error boxes for these three stars
are located outside the main sequence. Since  g modes are strongly damped after the main-sequence evolutionary stage,
it is rather improbable that these stars have already left the main sequence.

The evolutionary computations were performed using the Warsaw-New Jersey evolutionary
code \citep[e.\,g.][]{pamyatnykh1998}, with the OP opacity tables \citep{Seaton2005}
and the latest heavy element mixture of \citet{Asplund2009} (hereafter AGSS09).
We assumed the initial hydrogen abundance $X=0.7$,
the metallicities determined from observations (column 7, Table \ref{SPB_param})
and the equatorial rotational velocities equal to the measured projected value
(column 9, Table \ref{SPB_param}). Overshooting from convective core was not taken into account.

\begin{table*}
 \centering
 \begin{minipage}{\textwidth}
  \caption{The astrophysical parameters of the 31 selected SPB stars.
               Brightness in the second column is given in the Johnson $V$ passband.
               The parameters in the next eight columns have their usual meanings.
               Remarks are given in the last column.}
\label{SPB_param}
\centering
  \begin{tabular}{cccccccccccc}
  \hline
HD      & $V$     & SpT & $\log T_\mathrm{eff}$ & $\log L/\mathrm L_\odot$  & $\pi$ & $Z$ & $E(B-V)$ & $V_\mathrm{rot} \sin i$ & $M$ & rmk\\
       & (mag)  &      & (dex) & (dex)  & (mas) & (dex) & (mag) & (km s$^{-1}$) & (M$_\odot$) & \\

 \hline

1976      & 5.58$^1$  & B5IV$^1$      & 4.199$^2$    & 3.295      & 3.26$^3$  & 0.0165$^2$             & 0.091$^2$  & 143$^4$  & 6.2 & SB1$^{35}$\\
          &           &               & $\pm$0.005   & $\pm$0.176 & $\pm$0.63 & $^{+0.0008}_{-0.0007}$ &$\pm$0.004  & $\pm$9   &     & VB$^{36}$\\
3379      & 5.87$^1$  & B2.5IV$^1$    &4.301$^{12}$  & 3.397      & 3.14$^3$  & 0.0134$^*$             &0.070$^{16}$& 48$^{12}$& 7.1 &\\
          &           & B3V$^1$       & $\pm$0.022   & $\pm$0.101 & $\pm$0.30 &                        &            & $\pm$8   &     &\\
21071     & 6.08$^1$  & B5IV$^1$      & 4.164$^2$    & 2.444      & 6.27$^3$  & 0.0082$^2$             & 0.085$^2$  & 50$^{17}$& 3.7 &\\
          &           & B7V$^1$       & $\pm$0.007   & $\pm$0.076 & $\pm$0.40 & $^{+0.0016}_{-0.0014}$ & $\pm$0.005 &          &     &\\
24587     & 4.64$^4$  & B5V$^5$       & 4.161$^{12}$ & 2.626      & 8.65$^3$  & 0.0134$^*$             & 0$^{*}$   & 30$^{17}$& 4.2 & SB1$^{35}$\\
          &           &               & $\mp$0.030   & $\pm$0.060 & $\pm$0.30 &                        &            &          &     & \\
25558     & 5.33$^6$  & B5V$^7$       & 4.235$^2$    & 3.117      & 5.09$^3$  & 0.0096$^2$             & 0.112$^2$  & 28$^{12}$& 5.5 &\\
          &           &               & $\pm$0.008   & $\pm$0.071 & $\pm$0.32 & $^{+0.0019}_{-0.0016}$ & $\pm$0.010 & $\pm$2   &     & \\
26326     & 5.43$^8$  & B5IV$^5$      & 4.176$^2$    & 2.824      & 5.00$^3$  & 0.0103$^2$             & 0.004$^2$  & 17$^{12}$& 4.5 &\\
          &           &               & $\pm$0.006   & $\pm$0.070 & $\pm$0.28 & $^{+0.0018}_{-0.0015}$    & $\pm$0.004 & $\pm$1   &     & \\
28114     & 6.06$^1$  & B6IV$^1$      & 4.146$^{12}$ & 2.672      & 5.05$^3$  & 0.0134$^*$             &0.149$^{22}$& 21$^{12}$& 4.3 &\\
          &           &               & $\pm$0.031   & $\pm$0.099 & $\pm$0.49 &                        &            & $\pm$4   &     & \\
28475     & 6.79$^9$  & B5V$^{10}$    & 4.188$^{9}$  & 2.733      & 4.31$^3$  & 0.0134$^*$             & 0.240$^9$  & 30$^{29}$& 4.6 & SB2$^{35}$\\
          &           &               & $\pm$0.012   & $\pm$0.129 & $\pm$0.58 &                        & $\pm$0.010 &          &     & \\
34798     & 6.36$^1$  & B5IV/V$^1$    & 4.197$^2$    & 2.766      & 3.80$^3$  & 0.0115$^2$             & 0.038$^2$  & 30$^{17}$& 4.6 & VB$^{36}$\\
          &           & B2IV/V$^1$    & $\pm$0.006   & $\pm$0.119 & $\pm$0.47 & $^{+0.0020}_{-0.0017}$ & $\pm$0.004 & $\pm$9   &     &\\
          &           & B3V$^1$       &              &            &           &                        &            &          &     &\\
          &           & B5V$^1$       &              &            &           &                        &            &          &     &\\
37151     &7.38$^{11}$& B8V$^{12}$    & 4.108$^2$    & 1.895      & 5.20$^3$  & 0.0103$^2$             & 0.037$^2$  & 20$^{12}$& 2.8 &\\
          &           &               & $\pm$0.008   & $\pm$0.109 & $\pm$0.62 & $^{+0.0010}_{-0.0009}$ & $\pm$0.005 & $\pm$2   &     & \\
45284     &7.38$^{13}$& B8$^{13}$     & 4.176$^{12}$ & 2.476      & 3.02$^3$  & 0.0134$^*$             & 0$^{*}$   & 52$^{12}$& 4.1 &\\
          &           &               & $\pm$0.014   & $\pm$0.166 & $\pm$0.55 &                        &            & $\pm$5   &     & \\
53921     & 5.61$^1$  & B9IV$^1$      & 4.137$^{19}$ & 2.388      & 6.61$^3$  & 0.0134$^*$             & 0$^{*}$   & 17$^{31}$& 3.7 & VB$^{36}$\\
          &           & B9III$^1$     & $\pm$0.002   & $\pm$0.056 & $\pm$0.32 &                        &            & $\pm$10  &     &\\
74195     &3.62$^1$   & B3/5V$^1$     & 4.217$^{12}$ & 3.392      & 6.61$^3$  & 0.0100$^2$             & 0.006$^2$  & 18$^{12}$& 6.1 &\\
          &           & B3IV$^1$      & $\pm$0.026   & $\pm$0.069 & $\pm$0.35 & $\pm$0.0007            & $\pm$0.004 & $\pm$2   &     &\\
          &           & B3III$^1$     &              &            &           &                        &            &          &     &\\
74560     & 4.84$^1$  & B3IV$^1$      & 4.227$^2$    & 2.919      & 6.73$^3$  & 0.0177$^2$             & 0.004$^2$  & 45$^{30}$& 5.4 &SB1$^{35}$ \\
          &           & B3V$^1$       & $\pm$0.006   & $\pm$0.048 & $\pm$0.17 & $^{+0.0030}_{-0.0026}$ & $\pm$0.004 &          &     &VB$^{36}$\\
          &           & B4IV$^1$      &              &            &           &                        &            &          &     &\\
          &           & B5IV$^1$      &              &            &           &                        &            &          &     &\\
85953     & 5.93$^1$  & B2III$^1$     & 4.301$^{12}$ & 3.973      & 1.42      & 0.0134$^*$             & 0$^{*}$   & 30$^{12}$& 8.8 &\\
          &           & B2/3III$^1$   & $\pm$0.011   & $\pm$0.140 & $\pm$0.21 &                        &            & $\pm$1   &     &\\
          &           & B2V$^1$       &              &            &           &                        &            &          &     &\\
92287     & 5.90$^1$  & B3IV$^1$      & 4.215$^{19}$ & 3.178      & 3.24$^3$  & 0.0134$^*$             &0.070$^{25}$& 41$^{20}$& 5.8 & SB1$^{35}$\\
          &           & B3III$^1$     & $\pm$0.004   & $\pm$0.076 & $\pm$0.21 &                        &            & $\pm$16  &     &\\

121190    &5.65$^{37}$& B8V$^{38}$    & 4.082$^{40}$ & 1.992      & 9.24$^3$  & 0.0134$^*$             &0.010$^{38}$&118$^{40}$& 3.0 & \\
          &           & B9V$^{39}$    & $\pm$0.010   & 0.040      & $\pm$0.28 &                        &            & $\pm$3   &     & \\

123515    & 5.98$^1$  & B8V$^1$       & 4.079$^{19}$ & 2.574      & 3.98$^3$  & 0.0134$^*$             & 0$^{*}$    & 24$^{32}$& 3.7 & SB2$^{35}$\\
          &           & B8IV$^1$      & $\pm$0.002   & $\pm$0.090 & $\pm$0.39 &                        &            &          &     &VB$^{36}$\\
          &           & B9IV$^1$      &              &            &           &                        &            &          &     &\\
138764    & 5.16$^1$  & B8V$^1$       & 4.161$^{12}$ & 2.541      & 8.17$^3$  & 0.0134$^*$             &0.060$^{26}$& 21$^{12}$& 4.1 &\\
          &           & B6IV$^1$      & $\pm$0.030   & $\pm$0.081 & $\pm$0.58 &                        &            & $\pm$2   &     &\\
          &           & B7IV$^1$      &              &            &           &                        &            &          &     &\\
140873    & 5.39$^1$  & B8IV/V$^1$    & 4.144$^{19}$ & 2.583      & 7.25$^3$  & 0.0134$^*$             &0.116$^{24}$& 80$^{29}$& 4.1 & SB2$^{35}$ \\
          &           & B8IV$^1$      & $\pm$0.003   & $\pm$0.063 & $\pm$0.31 &                        &            &          &     &VB$^{36}$\\
          &           & B8III$^1$     &              &            &           &                        &            &          &     &\\
143309    &9.28$^{14}$& B8Ib/II$^{15}$& 4.147$^{20}$ & 2.557      &1.30$^{21}$& 0.0134$^*$             &0.140$^{14}$& 10$^{33}$& 4.1 &\\
          &           & B9Ib/II$^{15}$& $\pm$0.020   & $\pm$0.052 &           &                        &            &          &     &\\
160124    & 7.17$^1$  & B3IV/V$^1$    & 4.171$^{20}$ & 3.026      & 2.16$^3$  & 0.0134$^*$             &0.150$^{27}$& 8$^{20}$ & 5.2 &\\
          &           & B3IV$^1$      & $\pm$0.020   & $\pm$0.239 & $\pm$0.58 &                        &            & $\pm$4   &     &\\
          &           & B5V$^1$       &              &            &           &                        &            &          &     &\\
160762    & 3.80$^1$  & B3IV$^1$      & 4.290$^{12}$ & 3.396      & 7.17$^3$  & 0.0134$^*$             &0.000$^{28}$& 6$^{34}$ & 6.9 & SB1$^{35}$ \\
          &           & B3V$^1$       & $\pm$0.011   & $\pm$0.059 & $\pm$0.13 &                        &            & $\pm$1   &     &VB$^{36}$\\
          &           & B3.5III$^1$   &              &            &           &                        &            &          &     &\\
177863    & 6.30$^1$  & B8V$^1$       & 4.153$^2$    & 2.520      & 5.38$^3$  & 0.0132$^2$             & 0.126$^2$  & 60$^{17}$& 4.0 & SB1$^{35}$ \\
          &           & B8III$^1$     & $\pm$0.012   & $\pm$0.137 & $\pm$0.75 & $^{+0.0030}_{-0.0024}$ & $\pm$0.029 &          &     &VB$^{36}$\\

\end{tabular}
\end{minipage}
\end{table*}

\begin{table*}
 \centering
 \begin{minipage}{\textwidth}
  \contcaption{}
\centering
  \begin{tabular}{cccccccccccc}
  \hline
HD      & $V$     & SpT & $\log T_\mathrm{eff}$ & $\log L/\mathrm L_\odot$  & $\pi$ & $Z$ & $E(B-V)$ & $V_\mathrm{rot} \sin i$ & $M$ & rmk\\
       & (mag)  &      & (dex) & (dex)  & (mas) & (dex) & (mag) & (km s$^{-1}$) & (M$_\odot$) & \\

 \hline
179588    & 6,75$^1$  & B9IV$^1$      & 4.048$^{18}$ & 2.307      & 4.12$^3$  & 0.0134$^{18}$          &0.029$^{23}$& 35$^{17}$& 3.4 & SB1$^{35}$ \\
          &           & B8V$^1$       & $\pm$0.009   & $\pm$0.180 & $\pm$0.84 & $^{+0.0034}_{-0.0027}$ &            &          &     &VB$^{36}$\\
181558    & 6.26$^6$  & B5III$^5$     & 4.174$^2$    & 2.749      & 4.17$^3$  & 0.0135$^2$             & 0.077$^2$  & 20$^{17}$& 4.6 & VB$^{36}$\\
          &           &               & $\pm$0.006   & $\pm$0.094 & $\pm$0.38 & $^{+0.0027}_{-0.0022}$ & $\pm$0.005 &          &     & \\
182255    & 5.19$^1$  & B6III$^1$     & 4.148$^2$    & 2.414      & 8.30$^3$  & 0.0100$^2$             & 0.002$^2$  & 25$^{17}$& 3.7 &\\
          &           &               & $\pm$0.004   & $\pm$0.081 & $\pm$0.60 & $^{+0.0017}_{-0.0015}$ & $\pm$0.001 & $\pm$9   &     & \\
191295    & 7.16$^8$  & B7III$^8$     & 4.114$^{12}$ & 2.435      & 2.98$^3$  & 0.0134$^*$             & 0$^{*}$   & 16$^{12}$& 3.7 &\\
          &           &               & $\pm$0.033   & $\pm$0.141 & $\pm$0.47 &                        &            & $\pm$1   &     & \\
206540    & 6.05$^1$  & B5IV$^1$      & 4.130$^{12}$ & 2.586      & 4.66$^3$  & 0.0134$^*$             &0.048$^{24}$& 15$^{12}$& 4.1 &\\
          &           & B7III$^1$     & $\pm$0.032   & $\pm$0.078 & $\pm$0.37 &                        &            & $\pm$2   &     &\\
208057    & 5.08$^8$  & B3V$^8$       & 4.229$^2$    & 3.070      & 5.16$^3$  & 0.0110$^{2}$           & 0.020$^2$  &110$^{17}$& 5.5 &\\
          &           &               & $\pm$0.005   & $\pm$0.058 & $\pm$0.24 & $\pm$0.0005            & $\pm$0.004 & $\pm$16  &     & \\
215573    & 5.34$^1$  & B5/7V$^1$     & 4.153$^2$    & 2.591      & 6.62$^3$  & 0.0110$^2$             & 0.017$^2$  & 8$^{12}$ & 4.0 &\\
          &           & B6IV$^1$      & $\pm$0.003   & $\pm$0.057 & $\pm$0.19 & $^{+0.0008}_{-0.0007}$ & $\pm$0.003 & $\pm$1   &     &\\
\hline
\footnotetext{
 (1)  \citet{Reed2005};         (2) \citet{Niemczura2003};   (3) \citet{vanLeeuwen2007};
 (4)  \citet*{Huang2010};        (4) \citet{Clausen1997};     (5) \citet{Houk1988};
 (6)  \citet{Ducati2002};       (7) \citet{Cucchiaro1980};   (8) \citet*{Hohle2010};
 (9)  \citet{Fitzpatrick2007};  (10) \citet{Cowley1969};     (11) \citet{Warren1977A};
 (12) \citet{Lefever2010};      (13) \citet{Koen2002};       (14) \citet{Kilambi1975};
 (15) \citet{Houk1975};         (16) \citet{Gudennavar2012}; (17) \citet*{Abt2002};
 (18) \citet{Wu2011AA};         (19) \citet{Briquet2007};    (20) \citet{Hubrig2006};
 (21) \citet{Kharchenko2009};   (22) \citet{Vogt1998};       (23) \citet{Grillo1992};
 (24) \citet{Tobin1985};        (25) \citet{Turner1973};
 (26) \citet{Wegner2003};       (27) \citet{Vleeming1974};   (28) \citet{Nieva2012};
 (29) \citet*{Strom2005};        (30) \citet{Aerts1999872A};  (31) \citet{Hubrig2009};
 (32) \citet{Zorec2012};        (33) \citet{Grosso2011};     (34) \citet{Bailey2013};
 (35) \citet{Pourbaix2004};     (36) \citet{Mason2001};      (37) \citet{Kharchenko_Roeser2009};
 (38) \citet{Whittet1980};      (39) \citet{Houk1978};       (40) \citet{Aerts_Kolenberg2005};
 (*) $Z_\odot=0.0134$ assumed.
}

\end{tabular}
\end{minipage}
\end{table*}

We assumed the minimum value of the rotational velocity and the zero value
for convective core overshooting because their real values are not known.
It has to be emphasized that these parameters (like metallicity, initial abundance of hydrogen, etc.)
can affect the position and extent of evolutionary tracks on the H--R diagram
as well as the parameters estimated from the fit.
For instance, if we assume the often-used value of overshooting, $\alpha_\mathrm{ov}=0.2\,H_p$,
luminosities and effective temperatures from the centres of the error boxes
of HD\,74195, HD\,85953 and HD\,123515 are in agreement with the main-sequence
stages of evolution in contrast to the case without overshooting.
From these models and for the central values of
$\log T_\mathrm{eff}$ and $\log L/\mathrm L_\odot$, we obtained masses of 6.0\,M$_\odot$ for HD\,74195,
9.1\,M$_\odot$ for HD\,85953 and 3.8M\,$_\odot$ for HD\,123515
compared to the values 6.1\,M$_\odot$, 8.8\,M$_\odot$ and 3.7\,M$_\odot$
from the left bottom edges with neglected overshooting, respectively.
The models with overshooting give also different maximum values of the rotational frequencies,
$\nu_\mathrm{rot}^\mathrm{crit}=0.7623$ d$^{-1}$ for HD\,74195,
$\nu_\mathrm{rot}^\mathrm{crit}=0.6147$ d$^{-1}$ for HD\,85953 and
$\nu_\mathrm{rot}^\mathrm{crit}=0.9576$ d$^{-1}$ for HD\,123515
compared to
$\nu_\mathrm{rot}^\mathrm{crit}=1.0388$ d$^{-1}$,
$\nu_\mathrm{rot}^\mathrm{crit}=0.8281$ d$^{-1}$
and $\nu_\mathrm{rot}^\mathrm{crit}=1.1284$ d$^{-1}$
obtained from models from left-bottom edges with neglected overshooting, respectively.
Therefore, based on the differences between the cases with and without overshooting,
we can estimate the error in mass in Table\,\ref{SPB_param} to be at least
a few tenths of a solar mass, and the error in the critical rotation frequency
in Table\,\ref{SPB_puls} to be at least 0.25 d$^{-1}$.

\begin{figure*}
\includegraphics[angle=0, width=2\columnwidth]{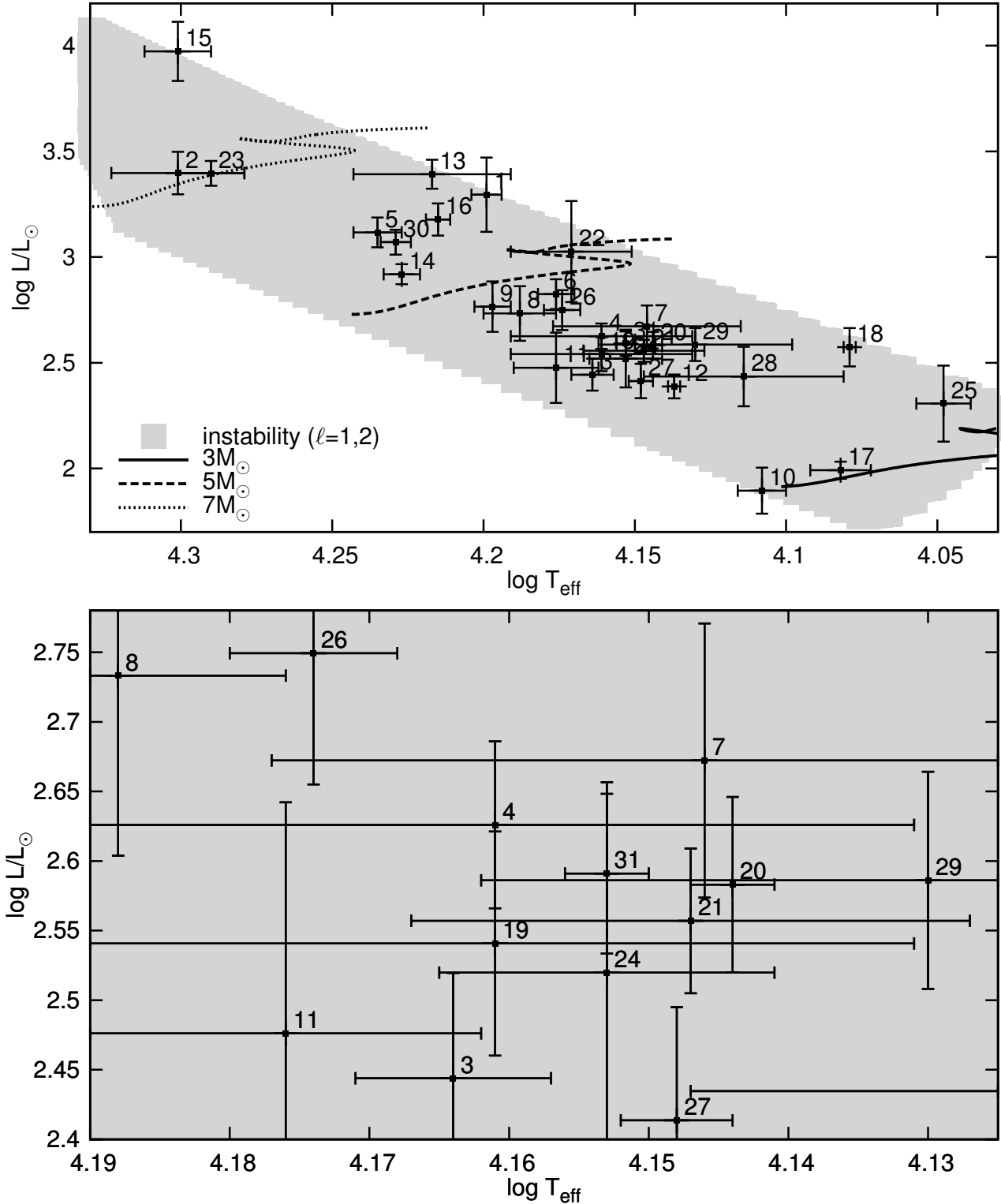}
\caption{The programme stars (cf.Table\,\ref{SPB_param}) in the H--R diagram.
           A zoom-in on the central, densely populated area around $\log T_\mathrm{eff}=4.16$
           and $\log L/\mathrm L_\odot=2.6$ is shown in the bottom panel for clarity.
         The SPB instability strip
         for the dipole and quadrupole modes is shown as a grey area.
         Stability calculations were performed
         with the initial abundance of hydrogen $X=0.7$, metallicity $Z=0.015$,
         OP opacities, AGSS09 element mixture, for non-rotating models,
         and without taking into account overshooting from the convective core.
         There are shown the evolutionary tracks for the masses 3$\mathrm M_\odot$ (solid line), 5$\mathrm M_\odot$ (dashed line) and 7$\mathrm M_\odot$ (dotted line).
         The labels (1$-$31) indicate the following stars:
         1 - HD\,1976; 2 - HD\,3379; 3 - HD\,21071; 4 - HD\,24587;
         5 - HD\,25558; 6 - HD\,26326; 7 - HD\,28114; 8 - HD\,28475;
         9 - HD\,34798; 10 - HD\,37151; 11 - HD\,45284; 12 - HD\,53921;
         13 - HD\,74195; 14 - HD\,74560; 15 - HD\,85953; 16 - HD\,92287;
         17 - HD\,121190;
         18 - HD\,123515; 19 - HD\,138764; 20 - HD\,140873; 21 - HD\,143309;
         22 - HD\,160124; 23 - HD\,160762; 24 - HD\,177863; 25 - HD\,179588;
         26 - HD\,181558; 27 - HD\,182255; 28 - HD\,191295; 29 - HD\,206540;
         30 - HD\,208057; 31 - HD\,215573. \label{HR}}
\end{figure*}

In Fig.\,1, the programme stars are plotted in the H--R diagram. Also shown is
the SPB instability strip for the dipole and quadrupole modes. To determine
the extent and position of the instability strip,
we used models with the parameters as described earlier, but with the metallicity $Z=0.015$
and with the rotational velocity $V_\mathrm{rot}$=0 km s$^{-1}$.
The numbers are assigned to the individual stars.
{ It should be noted that due to various metallicities and rotational
velocities, the models used to estimate masses listed in Table\,\ref{SPB_param}
do not correspond strictly to the evolutionary tracks presented in Fig.\,1.}

\begin{table*}
 \centering
 \begin{minipage}{\textwidth}
  \caption{The frequencies and amplitudes in the $V$ Geneva filter of the selected SPB stars.
              Symbols p and s denote that the frequency is detected
	      in the photometric and spectroscopic observations, respectively.
              The two second last columns contain the minimum and critical values of the rotational frequency.}
\label{SPB_puls}
\centering
  \begin{tabular}{ccccccccc}
  \hline
      & \multicolumn{4}{c}{$\nu\,\,\left(\mathrm d^{-1}\right)$} &   &  & \\
  HD      &\multicolumn{4}{c}{$A_V$ (mag)}  &   $\nu_\mathrm{rot}^\mathrm{min}$ $\left(\mathrm d^{-1}\right)$ & $\nu_\mathrm{rot}^\mathrm{crit}$ $\left(\mathrm d^{-1}\right)$ &Ref    \\

 \hline


1976      & 0.39946(4)$^\mathrm p$  & 0.93895(4)$^\mathrm p$    & 1.20346(6)$^\mathrm p$    &                   & 0.4770 & 0.8138 &    p - (1)   \\
          & 6.9(6)          & 5.9(6)            & 4.0(6)            &                   &        &        &       \\
3379      & 1.82023(19)$^\mathrm p$ & 1.59418(22)$^\mathrm p$   &                   &                   & 0.2273 & 1.4643 &   p - (1)  \\
          & 3.8(6)          & 3.6(6)            &                   &                   &        &        &   \\
21071     &1.18843(1)$^\mathrm {ps}$& 1.14934(3)$^\mathrm p$    & 1.41968(7)$^\mathrm p$    & 0.95706(9)$^\mathrm p$    & 0.3768 & 2.1256 &  p - (1)\\
          & 18.5(6)         & 7.7(6)            & 3.8(6)            & 3.0(6)            &        &        &  s - (2) \\
24587     &1.1569(6)$^\mathrm {ps}$ &                   &                   &                   & 0.1814 & 1.6385 & ps - (3) \\
          & 6.6(8)          &                   &                   &                   &        &        &   \\
25558     &0.65265(2)$^\mathrm {ps}$& 1.93235(8)$^\mathrm p$    & 1.17913(10)$^\mathrm p$   &                   & 0.1351 & 1.3287 & p - (1) \\
          & 14.7(6)         & 3.8(7)            & 3.6(6)            &                   &        &        & s - (2) \\
26326     & 0.5338(8)$^\mathrm {ps}$& 0.1723(8)$^\mathrm {s}$   & 0.7629(10)$^\mathrm p$    &                   & 0.0875 & 1.3296 & ps - (3) \\
          & 8.3(1.0)        &       -           & 7.0(1.0)          &                   &        &        & \\
28114     & 0.48790(4)$^\mathrm p$  & 0.48666(8)$^\mathrm p$    &                   &                   & 0.1123 & 1.3671 & p - (1) \\
          & 14.4(9)         &  7.2(1.0)         &                   &                   &        &        & \\
28475     & 0.68369(7)$^\mathrm p$  & 0.40893(13)$^\mathrm p$   &                   &                   & 0.1810 & 1.6978 & p - (1) \\
          & 8.4(1.0)        & 5.3(1.0)          &                   &                   &        &        &  \\
34798     & 0.78350$^\mathrm p$     & 0.85231$^\mathrm p$       &0.97432$^\mathrm p$        &  0.67795$^\mathrm p$      & 0.1818 & 1.7114 & p - (2) \\
          & 12.6(7)         & 2.4(7)            & 5.9(7)            & 7.0(7)            &        &        &  \\
37151     & 1.243166$^\mathrm p$    & 1.180186$^\mathrm p$      & 1.042663$^\mathrm p$      & 1.105252$^\mathrm p$      & 0.2195 & 3.2770 & p - (4) \\
          & 9.5(4)          & 9.8(4)            &   8.7(4)          &  7.7(4)           &        &        & \\
          & 1.452063$^\mathrm p$    &                   &                   &                   &        &        &  \\
          & 1.9(4)          &                   &                   &                   &        &        &  \\
45284     & 1.23852$^\mathrm p$     &   1.12753$^\mathrm p$     &  1.50605$^\mathrm p$      &                   & 0.4006 & 2.3120 & p - (2) \\
          & 7.4(7)          & 10.5(8)           & 4.7(7)            &                   &        &        &  \\
53921     & 0.6054(6)$^\mathrm {ps}$&                   &                   &                   & 0.1209 & 1.9618 & ps - (3) \\
          & 5.3(7)          &                   &                   &                   &        &        &  \\

74195     &0.35475(9)$^\mathrm {ps}$&0.35033(9)$^\mathrm {ps}$  & 0.34630(9)$^\mathrm {ps}$ & 0.39864(9)$^\mathrm {ps}$ & 0.0710 & 1.0388 & ps - (3) \\
          &  13.2(4)        &  10.6(4)          & 3.5(4)            &  3.0(4)           &        &        & \\
74560     &0.644714(2)$^\mathrm {ps}$& 0.395771(10)$^\mathrm p$ & 0.447646(21)$^\mathrm p$  & 0.635982(33)$^\mathrm p$  & 0.2632 & 1.7639 & ps - (*) \\
          & 15.0(4)          &  4.2(4)          &  3.3(4)           &  2.2(4)           &        &        & \\
          &0.822845(43)$^\mathrm p$ & 1.610567(54)$^\mathrm p$  &                   &                   &        &        &  \\
          & 3.1(4)          & 2.6(4)            &                   &                   &        &        &   \\
85953     & 0.2663(6)$^\mathrm {ps}$& 0.2189(6)$^\mathrm p$     & 0.2353(7)$^\mathrm s$     &                   & 0.0903 & 0.8281 & ps - (3) \\
          & 8.5(7)          & 3.8(7)            &  -                &                   &        &        & \\
92287     &0.21480(7)$^\mathrm {ps}$&                   &                   &                   & 0.1681 & 1.0669 & ps - (3) \\
          &   10.2(6)       &                   &                   &                   &        &        & \\
121190    & 2.6831(4)$^\mathrm {ps}$& 2.6199(4)$^\mathrm {p}$  & 2.4713(7)$^\mathrm {p}$    &                   & 1.0273 & 2.3831 & ps - (4)\\
          & 4.8(7)          & 3.9(6)           & 2.6(8)             &                   &        &        & \\
123515    &0.68528(10)$^\mathrm {ps}$& 0.65929(10)$^\mathrm p$  & 0.72585(10)$^\mathrm {ps}$& 0.55198(10)$^\mathrm p$   & 0.1181 & 1.1284 & ps - (3) \\
          & 16.4(5)         & 11.3(5)           &  9.3(5)           &  5.2(5)           &        &        &  \\
138764    & 0.7944(9)$^\mathrm {ps}$& 0.6372(9)$^\mathrm s$     &                   &                   & 0.1395 & 1.8539 & ps - (3) \\
          & 17.7(1.1)       &   -               &                   &                   &        &        &  \\
140873    & 1.1515(8)$^\mathrm {ps}$&                   &                   &                   & 0.4690 & 1.5388 & ps - (3) \\
          &  15.4(1.6)      &                   &                   &                   &        &        &  \\
143309    & 0.59830(2)$^\mathrm p$  & 0.60213(2)$^\mathrm p$    & 0.71414(2)$^\mathrm p$    & 1.20066(2)$^\mathrm p$    & 0.0611 & 1.6277 & p - (2)  \\
          & 23.0(1.5)       & 17.6(1.5)         & 10.7(1.5)         & 6.7(1.4)          &        &        &  \\
          & 0.59133(2)$^\mathrm p$  &                   &                   &                   &        &        & \\
          & 7.6(1.4)        &                   &                   &                   &        &        &   \\
160124    & 0.52014$^\mathrm p$     &  0.52096$^\mathrm p$      &   0.52055$^\mathrm p$     & 0.52137$^\mathrm p$       & 0.0319 & 0.9692 & p - (2)  \\
          & 11.6(1.6)       & 10.1(1.4)         & 12.7(1.4)         & 11.0(1.5)         &        &        &  \\
          & 0.52197$^\mathrm p$     & 0.70259$^\mathrm p$       & 1.0422$^\mathrm p$        & 0.31366$^\mathrm p$       &        &        &  \\
          & 12.3(1.5)       & 7.6(1.3)          & 5.8(1.3)          & 5.7(1.3)          &        &        &  \\
160762    & 0.28671$^\mathrm p$     &                   &                   &                   & 0.0360 & 1.3462 & p - (2)  \\
          &  7.8(2.9)       &                   &                   &                   &        &        &  \\
177863    &0.84059(10)$^\mathrm {ps}$& 0.10108(10)$^\mathrm p$  &                   &                   & 0.3943 & 1.8074 & ps - (3) \\
          &  17.4(4)        & 2.8(4)            &                   &                   &        &        &  \\
179588    & 0.85654(4)$^\mathrm p$  &   2.04263(5)$^\mathrm p$  &  2.19989(7)$^\mathrm p$   &  1.83359(9)$^\mathrm p$   & 0.1815 & 1.1582 & p - (1)\\
          & 12.9(1.3)       & 10.5(1.3)         & 8.0(1.3)          & 5.9(1.4)          &        &        &  \\
181558    &0.80780(10)$^\mathrm {ps}$&                  &                   &                   & 0.1114 & 1.4997 & ps - (3) \\
          & 27.5(7)         &                   &                   &                   &        &        &  \\

\end{tabular}
\end{minipage}
\end{table*}

\begin{table*}
 \centering
 \begin{minipage}{\textwidth}
  \contcaption{}
\centering
  \begin{tabular}{ccccccccc}
  \hline
      & \multicolumn{4}{c}{$\nu\,\,\left(\mathrm d^{-1}\right)$} &   &  & \\
  HD      &\multicolumn{4}{c}{$A_V$ (mag)}  &   $\nu_\mathrm{rot}^\mathrm{min}$ $\left(\mathrm d^{-1}\right)$ & $\nu_\mathrm{rot}^\mathrm{crit}$ $\left(\mathrm d^{-1}\right)$ &Ref    \\

 \hline

182255    &0.97185(3)$^\mathrm {ps}$& 0.79225(4)$^\mathrm {ps}$ & 0.62526(6)$^\mathrm p$    & 1.12780(10)$^\mathrm p$   & 0.1815 & 2.0047 & p - (1) \\
          & 18.7(9)         & 12.7(9)           & 8.3(8)            & 5.8(8)            &        &        & s - (2) \\
          &1.02884(11)$^\mathrm p$  &                   &                   &                   &        &        & \\
          & 4.3(1.1)        &                   &                   &                   &        &        &  \\
191295    & 0.71908(6)$^\mathrm p$  & 0.49011(12)$^\mathrm p$   &  0.29302(17)$^\mathrm p$  &                   & 0.0972 & 1.5462 & p - (1) \\
          & 15.7(9)         & 8.6(1.0)          & 5.8(9)            &                   &        &        &  \\
206540    & 0.72002(4)$^\mathrm p$  &  0.62125(5)$^\mathrm p$   & 0.38271(8)$^\mathrm p$    &                   & 0.0822 & 1.3811 & p - (1) \\
          & 12.2(1.1)       & 8.6(1.1)          &  5.6(1.0)         &                   &        &        &  \\
208057    &0.89050(7)$^\mathrm {ps}$& 0.80213(10)$^\mathrm {ps}$& 2.47585(8)$^\mathrm p$    &                   & 0.5440 & 1.3741 & p - (1) \\
          & 8.4(1.4)        & 6.0(1.3)          & 7.4(1.5)          &                   &        &        & s - (2) \\
215573    & 0.5654(6)$^\mathrm {ps}$&                   &                   &                   & 0.0484 & 1.5973 & ps - (3) \\
          & 14.9(2.0)       &                   &                   &                   &        &        &  \\
		
\hline
\end{tabular}
\footnotetext{
 (1)  \citet{DeCat2007};         (2) \citet{DeCat2002ASPC_259_A};   (3) \citet{DeCat2002};
 (4) \citet{North1994}; (4) \citet{Aerts_Kolenberg2005}; (*) this paper.
}
\end{minipage}
\end{table*}

\section{Mode identification}

In order to make mode identification, we will use observations in the $UBV$ Geneva filters.
We include only these three filters because the response functions in the Geneva photometric system widely overlap
and all seven filters do not give independent information.
Therefore, we chose the $UBV$ filters which have non-overlapping response functions.
In Table\,\ref{SPB_puls}, we list all frequencies detected in the programme stars,
and the corresponding values of the amplitudes in the $V$ filter.
In the two second last columns, we give also the minimum value of the rotational frequency, resulting from $V_{\rm rot}\sin i$,
and the critical rotational frequency corresponding to the break-up velocity $V_{\rm rot}^{\rm crit}\approx \sqrt{\frac{GM}{R}}$.
The values of a mass $M$ and a radius $R$ were adopted for the central models of Table\,1.

In this section, we present the results of mode identification
for the frequency peaks observed in the selected SPB stars.
First, we consider the case when all effects of rotation are neglected.
Then, we give results in the framework of the traditional approximation
in which the effect of the Coriolis force on long-period g modes is included.

\subsection{Methods}

\subsubsection{The zero-rotation approximation}
The most common way to identify pulsation modes from photometry  is to compare the empirical amplitude ratios and phase differences
in different photometric passbands with their theoretical counterparts.
Usually, we do this by computing a discriminant \citep[e.\,g.][]{JDDPW2009}
\begin{equation}
\chi^2=\frac{1}{2\left(N-1\right)} \sum_{i=1}^{2\left(N-1\right)} \frac{\left( X_i^o-X_i^t \right)^2}{\sigma_i^2},
\end{equation}
where $X_i^o$ and $X_i^t$ are  the observational and theoretical photometric observables, respectively,
i.e., the amplitude ratios, $A_x/A_y$, or phase differences, $\phi_x-\phi_y$, in $x$, $y$ photometric
passbands. $N$ denotes the number of passbands and $\sigma_i$ are the observational errors
of the quantities $X_i^o$.

If all effects of rotation on stellar pulsation are neglected,
the photometric observables are independent of the inclination angle, $i$,
intrinsic mode amplitude, $\varepsilon$, and the azimuthal order, $m$.
Thus, the method allows us to determine only the mode degree. This is the main disadvantage
of disregarding the rotational effects.

In the mode identification process, we accept the degree $\ell$ if the corresponding value of $\chi^2$
is below some specified level. We choose this level in such a way that the probability
of rejecting a correct solution is less than 20 per cent. 
In the case of three passbands, it is about $\chi^2\approx 1.5$.
We note, however, that it is quite a rough estimate, because we do not deal here with a strict
statistical approach. Moreover, due to an increase of the averaging over the stellar disc
with the increasing mode degree, we considered modes with  $\ell$ up to 6.

The formulae for the complex amplitude of the light variation in the $x$ passband in the framework
of the linear theory in the zero-rotation approximation were given, e.\,g. by \citet{JDD2002}
and are not repeated here.

To compute the theoretical values of the photometric amplitudes and phases, input
from models of stellar atmospheres and linear non-adiabatic pulsation calculations are needed.
The former include the flux derivatives over effective temperature and gravity
and the limb-darkening coefficients and their derivatives. While the latter, the so-called
non-adiabatic parameter $f$ defined as the ratio of the amplitude of the radiative
flux perturbation to the radial displacement at the photosphere level.
For details see, e.\,g. \citet*{JDD2003} and \citet{JDD2011}.
In what follows, we rely on the \citet{Kurucz2004} stellar atmosphere models.

A significant reduction in the number of modes to be considered comes from information about their instability.
First, we took into account only theoretically unstable modes, i.\,e., those with the instability parameter $\eta>0$.
However, because of uncertainties in the model and opacity data, this condition was slightly
relaxed, and ultimately we accepted the modes with $\eta>-0.05$.

\subsubsection{Including rotation}
Taking into account rotation in the mode identification demands a more sophisticated treatment
than that described in the preceding subsection.
We shall use the method proposed by \citet{JDD2008,JDD2015},
which includes the effects of rotation via the traditional approximation.
In this method, the observed complex amplitudes are compared with the corresponding theoretical values using the statistic:
\begin{equation}
\chi^2=\frac{1}{2N - M}  \sum^N_{i=1} w_i \left| \mathcal{A}_i -\varepsilon\mathcal{F}_i\right|^2,
\end{equation}
where $\mathcal A_i$ and  $\mathcal F_i$ are the  observed and theoretical complex amplitudes, respectively.
The values of  $w_i=1/\sigma_i^2$ are the statistical weights determined by the observational errors, $\sigma_i$.
$N$ denotes the number of passbands and $M$ is the number of  unknowns to be determined.
The value of the intrinsic mode amplitude, $\varepsilon$, can be obtained from observations by minimizing $\chi^2$:

\begin{equation}
  \varepsilon=\frac{\sum^N_{i=1} w_i \mathcal A_i   \mathcal F_i^*}{\sum^N_{i=1} w_i \left|\mathcal F_i \right|^2},
\end{equation}

The final expression for the mode discriminant is written as
\begin{equation}
 \chi^2  = \frac{1}{2N - M} \left( \sum^N_{i=1} w_i    \left| \mathcal A_i \right|^2 -
  \frac{\left( \sum^N_{i=1} w_i \left|\mathcal A_i \mathcal F_i^* \right|\right)^2}
  {\sum^N_{i=1} w_i \left| \mathcal F_i \right|^2} \right),
\end{equation}
where the asterisk denotes complex conjugate.

Now, the value of $\chi^2$  depends on both mode numbers ($\ell$, m), the inclination angle $i$ and
rotational velocity $V_\mathrm{rot}$.
To reduce the number of unknowns, the projected rotational velocity is kept constant
i.\,e., $V_\mathrm{rot}\sin i= \mathrm{const}$.
Thus, in principle, the method allows deriving simultaneously a value of the rotational velocity, $V_\mathrm{rot}$.

The theoretical amplitude ${\mathcal F}_i$ is calculated with the effects of rotation taken into account.
The expression for ${\mathcal F}_i$ can be found in \citet{JDD2007, JDD2015} and need not be repeated here.
Similarly to the zero-rotation case, two inputs are needed. We used the same atmosphere
models as before, but pulsational calculations which give us the $f$-parameter
are performed in the framework of the traditional approximation in which Coriolis force is included.

One of the assumptions of the traditional approximation is that centrifugal distortion of a star
is negligible. Therefore, we considered { the angular frequency of rotation}
up to $\Omega^2 \lesssim 0.5 \Omega_{\rm crit}^2$, where $\Omega_{\rm crit}$ is a critical value.
This condition corresponds approximately to the limit of the half of the critical equatorial velocity, $V_\mathrm{rot}^{\rm crit}$
(assuming that at $\Omega=\Omega_\mathrm{crit}$
we have $R_\mathrm{pole}=2/3 R_\mathrm{eq}$, where $R_\mathrm{pole}$ and $R_\mathrm{eq}$ are
the pole and equatorial radius, respectively).

As previously, only modes with $\ell \le 6$ are assumed to be visible in the ground-based photometry.
However, this time we are dealing with the rotational splitting 2$\ell$+1, so we have to consider
all modes with $| m| \le \ell$.
In addition, we include mixed-Rossby modes, which may be excited and visible in the light variations once rotation is fast enough.
They are retrograde with $m =-\ell$ and are sometimes called r modes \citep{Lee2006, JDD2007}.
Again only modes with $\eta>-0.05$ were considered.
The maximum accepted value of the $\chi^2$ is specified in the same way as in the case without rotation.
Moreover, \citet{JDD2008, JDD2015} showed that intrinsic mode amplitude in the SPB stars should not be greater than 0.01.
Therefore, we rejected all identifications with $|\varepsilon|>0.01$.

\begin{figure}
\centering
\includegraphics[angle=-90, width=\columnwidth]{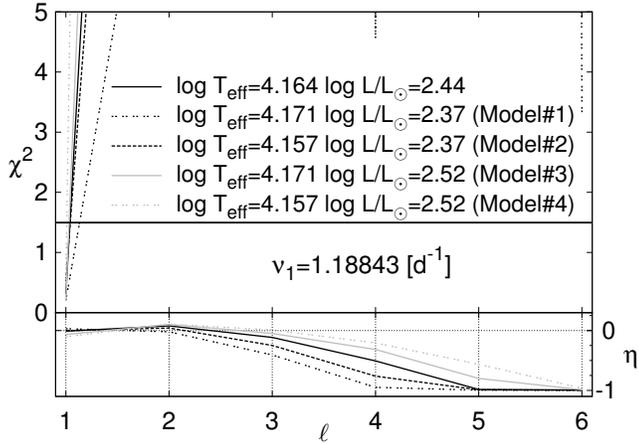}
\caption{The discriminant $\chi^2$ as a function of the mode degree $\ell$,
         for the dominant frequency of HD\,21071, calculated in the zero-rotation
		 approximation for models from the centre (the solid line) and four edges
         (the other lines) of the error box. The black horizontal line indicates our confidence level ($\chi^2$=1.5).
		 In the lower panel, there is shown the corresponding value of the instability parameter, $\eta$.\label{chi_0_rot}}
\end{figure}

If a given mode is also seen in the radial velocity variations, one can use
the information about the mode geometry contained in the amplitude of this variations.
Only modes which reproduce the amplitude of the radial velocity variations, $A_{V_{\rm rad}}$,
within the observational error are considered as responsible for a given frequency peak.
The value of $A_{V_{\rm rad}}$ was calculated with the empirical value of $\varepsilon$ obtained from photometry (see equation 3).
In the case of a star in which some frequencies are visible in, both, photometry and spectroscopy
and some only in the photometry, we adopted a fairly reasonable assumption that the radial velocity amplitudes of
modes invisible in spectroscopy should be smaller than the smallest observed amplitude.
Otherwise, such frequencies would be detected in spectroscopy.

\subsection{Working example: HD\,21071}
Our working example is HD\,21071 (V576 Per), the SPB star with four frequency peaks observed in photometry
(see Table\,\ref{SPB_puls}). The dominant frequency was also detected in the spectroscopic data.
As for the other stars, identification of pulsation modes in the case of HD 21071 was made
using the evolutionary models from which the masses given in the 10th column of Table\, \ref{SPB_param}
were estimated.

\begin{figure*}
\centering
\includegraphics[angle=-90, width=2\columnwidth]{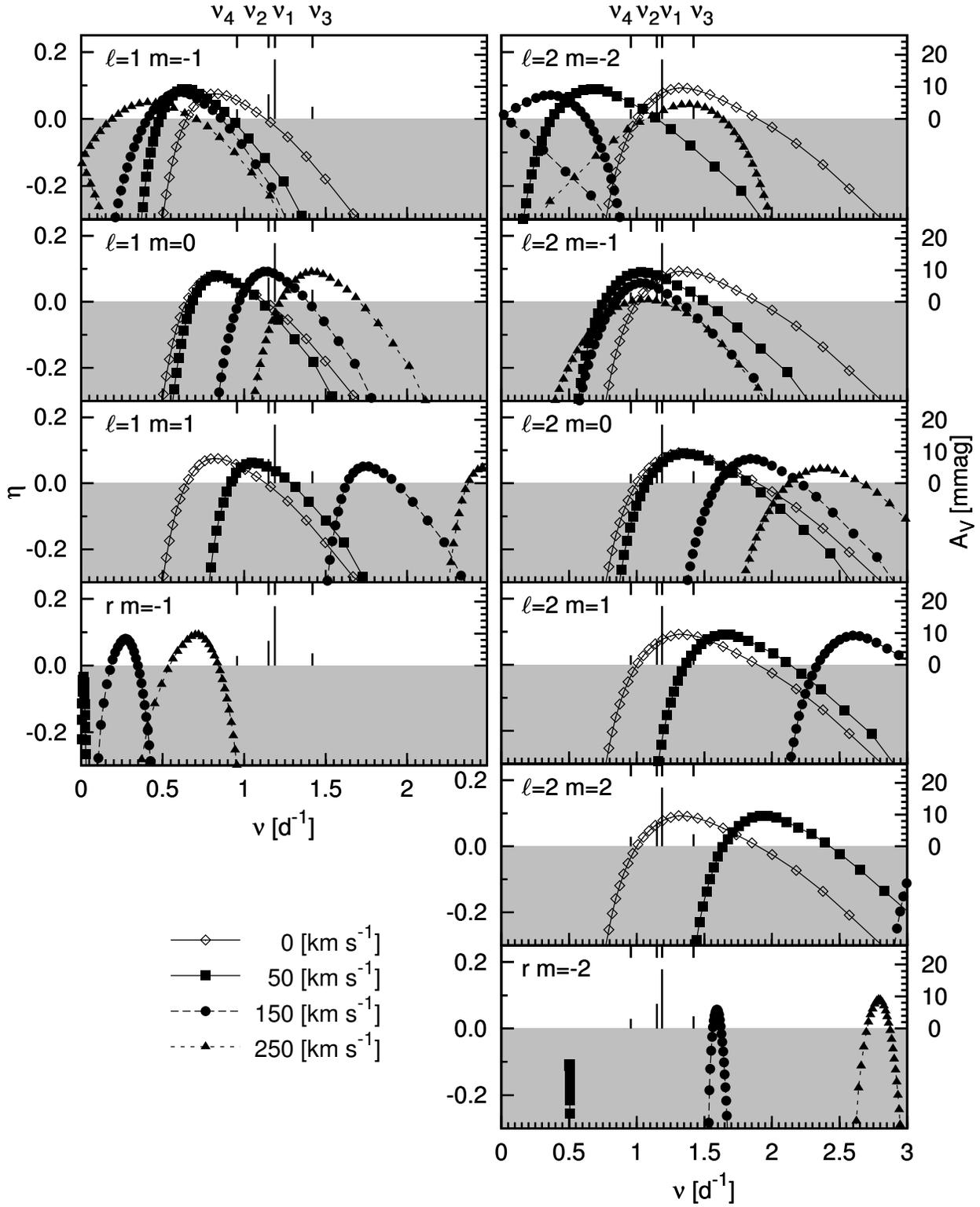}
\caption{The instability parameter for the dipole, quadrupole and r modes with $m=-1$, --2 in the central model of HD\,21071
         ($\log T_\mathrm{eff}=4.164$; $\log L/\mathrm L_\odot=2.444$; $M=3.69\mathrm M_\odot$).
         Four values of the rotational velocity were considered:
         0 km s$^{-1}$ (open diamonds), 50 km s$^{-1}$ (filled squares),
         150 km s$^{-1}$ (filled circles) and 250 km s$^{-1}$ (filled triangles).
         The observed Geneva $V$-filter amplitudes are plotted with the vertical solid lines.
		 \label{freq_eta_l1ApJS}}
\end{figure*}

\begin{figure*}
\centering
\includegraphics[angle=-90, width=2\columnwidth]{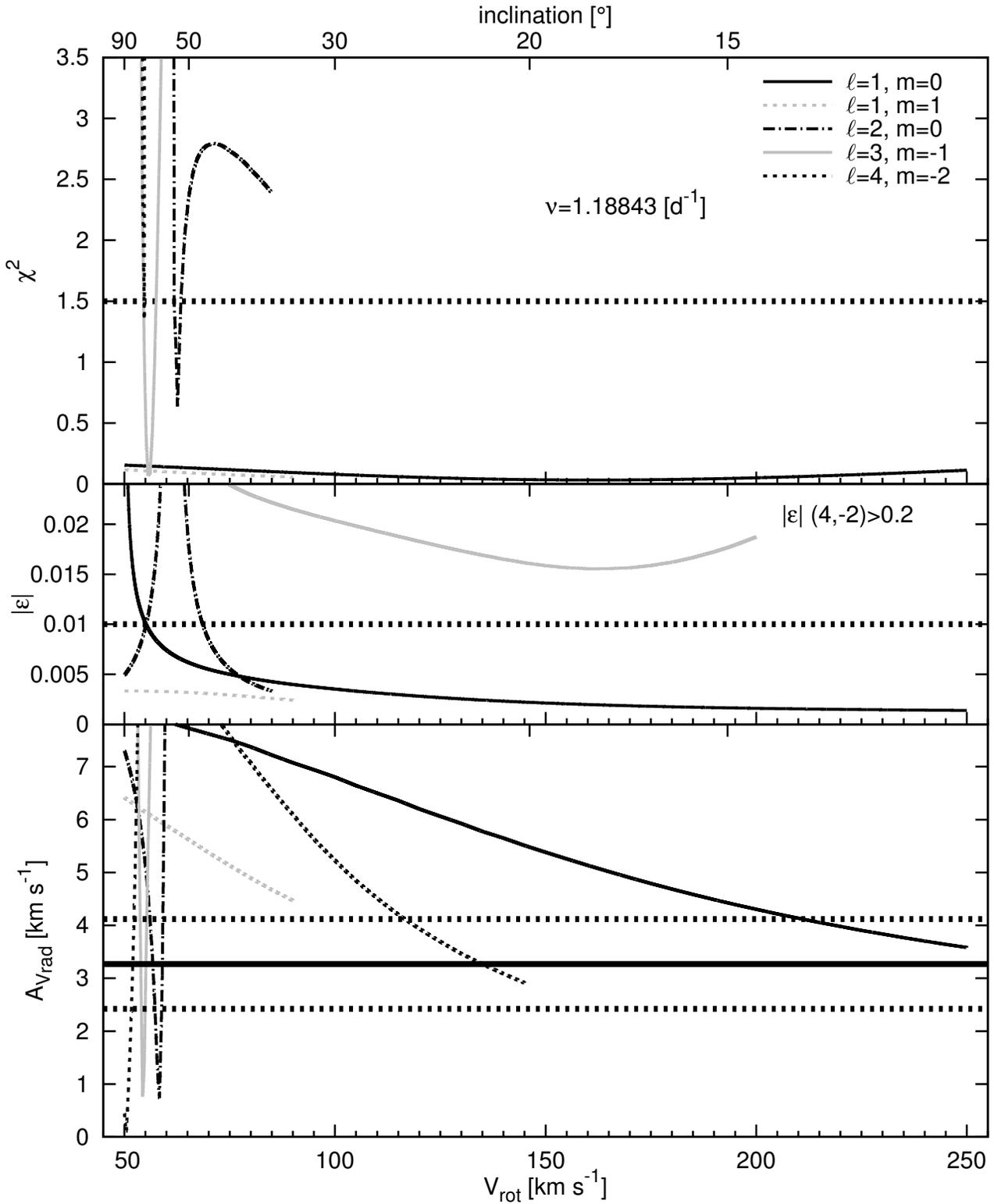}
\caption{Top panel: the value of $\chi^2$ as a function of the rotational velocity for the
         dominant frequency $\nu_1$=1.18843 d$^{-1}$ of HD\,21071, calculated using
         the traditional approximation.
         The stellar parameters of the model from the centre of the error box were assumed.
         Horizontal dotted line indicates the 80 per cent confidence level.\newline
         Middle panel: the value of the intrinsic mode amplitude, $\varepsilon$,
         as a function of the rotational velocity for the dominant
         frequency of HD\,21071.
         Only modes presented in the top panel
         are shown. The horizontal dotted line marks the adopted
         maximum of $|\varepsilon|$.\newline
         Bottom panel: the theoretical value of the radial velocity amplitude as
         a function of the rotational velocity for the dominant frequency observed in HD\,21071.
         Only modes presented in the top panel
         are shown. The observed range of the radial velocity amplitude for $\nu_1$ is marked with the
         horizontal lines.\label{chi_rot}}
\end{figure*}

\begin{table*}
 \centering
 \begin{minipage}{\textwidth}
  \caption{The results of mode identification of the frequency peaks observed in HD\,21071 from the four approaches described in the text. The adopted model parameters are: $\log T_\mathrm{eff}=4.164$, $\log L/\mathrm L_\odot=2.444$, $M=3.69\mathrm M_\odot$. The values of the rotational velocity and the rotational frequency are in km s$^{-1}$ and d$^{-1}$, respectively.
  Column {\it a}  $-$ solutions which satisfy the $\chi^2$ condition;  column {\it b} $-$  solutions
which
satisfy additionally the $\varepsilon$ condition;
column {\it c} $-$ solutions which satisfy the conditions $\chi^2$, $\varepsilon$,
and  A$_{Vr}$;
column {\it d} $-$ solutions which satisfy the conditions $\chi^2$, $\varepsilon$,
A$_{Vr}$,
and the condition on the common range of $V_\mathrm{rot}$.}
\label{mic_tab_HD21071}
\centering
  \begin{tabular}{llllll}
  \hline
        & $\ell$  &\multicolumn{3}{c}{$\left(\ell,m\right) ~\left[V_\mathrm{rot}^\mathrm{min};V_\mathrm{rot}^\mathrm{max}\right]$} &\multicolumn{1}{c}{$\left(\ell,m\right) ~\left[V_\mathrm{rot}^\mathrm{min};V_\mathrm{rot}^\mathrm{max}\right] ~\left\{\nu_\mathrm{rot}^\mathrm{min};\mathrm \nu_\mathrm{rot}^\mathrm{max} \right\}$}  \\
 $\nu$ (d$^{-1})$     & {$V_\mathrm{rot}$=0}   & ~~~~~~\emph{a}      &   ~~~~~~\emph b &  ~~~~~~\emph c &   ~~~~~~~~~~~~~~~~~~\emph d  \\

 \hline

        1.18843   & 1         & (1,0) [50;250]                  & (1,0) [55;250]  & (1,0) [211;250]  & (1,0) [211;212] \{1.5900;1.5976\}\\
                  &           & (1,1) [50;90]                   & (1,1) [50;90]   &                  &                 \\
                  &           & (2,0) [62;63]                   &                 &                  &      \\
                  &           & (3,-1) [55;58]                  &                 &                  &      \\
                  &           & (4,-2) [55;55]                  &                 &                  &      \\
\hline
        1.14934   & 1         & (1,0) [50;235]                  & (1,0) [51;235]  & (1,0) [51;235]   & (1,0) [211;212] \{1.5900;1.5976\}         \\
                  &           & (1,1) [50;85]                   & (1,1) [50;85]   & (1,1) [50;85]    &                \\
                  &           & (2,0) [50;60]                   & (2,0) [50;59]   & (2,0) [50;59]    &                \\
                  &           & (2,0) [62;75]                   & (2,0) [64;75]   &                  &      \\
                  &           & (3,-1) [54;66]                  &                 &                  &      \\
                  &           & (4,-2) [54;55]                  &                 &                  &   \\
\hline
        1.41968   & 1         & (1,0) [135;255]                 & (1,0) [135;255] & (1,0) [135;255]  &  (1,0) [211;212]  \{1.5900;1.5976\}        \\
                  &           & (1,1) [60;125]                  & (1,1) [60;125]  & (1,1) [60;125]   &              \\
                  &           & (2,0) [50;60]                    & (2,0) [50;59]   & (2,0) [50;59]    &            \\
                  &           & (2,0) [62;125]                  & (2,0) [64;125]  & (2,0) [69;125]   &               \\
                  &           & (2,-1) [50;50]                  & (2,1) [50;60]   & (2,1) [50;60]    &                 \\
                  &           & (2,1) [50;60]                   & (3,0) [62;65]   & (3,0) [62;65]    &             \\
                  &           & (3,0) [50;50]                   & (4,-1) [58;66]  & (4,-1) [58;66]   &                \\
                  &           & (3,0) [62;65]                   & (4,-1) [128;175]& (4,-1) [135;175] &             \\
                  &           & (3,-1) [52;63]                  &                 &                  &  \\
                  &           & (3,-2) [50;50]                  &                 &                  &   \\
                  &           & (4,-1) [50;50]                  &                 &                  &      \\
                  &           & (4,-1) [58;81]                  &                 &                  &      \\
                  &           & (4,-1) [128;175]              &                 &                  &      \\
                  &           & (4,-2) [54;55]                  &                 &                  &      \\
                  &           & (4,-3) [50;50]                  &                 &                  &      \\
                  &           & (5,-2) [50;50]                  &                 &                  &      \\
                  &           & (5,-2) [62;63]                  &                 &                  &      \\
                  &           & (5,-3) [53;53]                  &                 &                  &      \\
                  &           & (5,-4) [50;50]                  &                 &                  &      \\
                  &           & (6,-5) [50;50]                  &                 &                  &      \\
\hline

        0.95706   & 1, 2      & (1,0) [50;155]                  & (1,0) [50;155]  & (1,0) [50;155]   &  (2,-1) [211;212]  \{1.5900;1.5976\}       \\
                  &           & (1,-1) [50;105]                & (1,-1)[50;105]  & (1,-1) [50;105]  &                     \\
                  &           & (1,1) [50;60]                  & (1,1) [50;60]   & (1,1) [50;60]    &                 \\
                  &           & (2,-1) [50;212]               & (2,-1) [50;212] & (2,-1) [50;212]  &                     \\
                  &           & (2,-2) [50;57]                 & (2,-2) [50;57]  & (2,-2) [50;56]   &                 \\
                  &           & (2,-2) [180;196]             & (2,-2) [180;189]& (2,-2) [180;189] &                 \\
                  &           & (3,-2) [50;50]                 &                 &                  &      \\
                  &           & (4,-2) [55;56]                 &                 &                  &      \\
                  &           & (4,-3) [50;50]                 &                 &                  &      \\
                  &           & (5,-4) [50;50]                 &                 &                  &      \\
                  &           & (6,-5) [50;50]                 &                 &                  &      \\
\hline
\end{tabular}
\end{minipage}
\end{table*}

We started our analysis with the zero-rotation approximation. In Fig.\,\ref{chi_0_rot}, we show
the value of the discriminant $\chi^2$ for the dominant frequency observed in HD\,21071 as a function of the mode degree, $\ell$.
The models corresponding to the centre and four edges of the error box were considered.
As one can see, a deep minimum at $\ell=1$ is present regardless of the adopted model parameters.
Thus, the frequency $\nu_1$ is undoubtedly that of a dipole mode. The value of the instability parameter, $\eta$, for the $\ell=1$ mode
is close to 0 and therefore acceptable.
The results of mode identification in the zero-rotation approximation
for other frequencies are given in the second column of Table \ref{mic_tab_HD21071}.
The results for the other frequencies are usually model independent with the exception of $\nu_2$ and $\nu_3$
for which we obtained $\ell=1$ or 2 if parameters of Model\#1 or Model\#2 were assumed.
The effect of core overshooting on identification of $\ell$ was tested using Model\#5, i.e.,
the central model with $\alpha_\mathrm{ov}=0.2\,H_p$. This result is not shown in Fig.\,\ref{chi_0_rot}
because $\chi^2$ for this model almost coincides with $\chi^2$ for the central model
without overshooting and they are hardly distinguishable. Results for other frequencies are also
compatible with those from the central model without core overshooting.

Mode identification with the effects of rotation taken into account consists of
several steps. Calculations were performed in the range of rotational velocities
from 50  to 250 km s$^{-1}$ corresponding to the observed
value of $V_\mathrm{rot} \sin i$ and about a half of the
critical rotational velocity for HD\,21071, respectively.
First, we rejected all modes with $\eta<-0.05$.
In Fig.\,\ref{freq_eta_l1ApJS}, we show an example of the instability parameter for the dipole, quadrupole and r modes with $m=-1$, --2
for the three values of the rotational velocity, $V_\mathrm{rot} =50$, 150, 250 km s$^{-1}$, as well as for the zero-rotation case.
A model of HD\,21071 with the central values of the error box, $\log T_\mathrm{eff}=4.164$, $\log L/\mathrm L_\odot=2.444$, $M=3.69\mathrm M_\odot$,
was assumed.
In Fig.\,3, we mark also frequencies and the Geneva $V$ amplitudes of the four peaks detected in HD\,21071.
Relying only on the instability information, one can see that the observed frequencies can be associated
with some modes at certain values of the rotational velocities.

The frequency $\nu_1$ can be associated with the modes (1, 0) or (2,$-$1) in
the whole range of the rotational velocities. The modes (1, +1) or (2, 0) are possible
only for the lowest values of $V_\mathrm{rot}$, and (r, $-$2) for a narrow range of $V_\mathrm{rot}$ around 120 km s$^{-1}$.
The peak $\nu_1$ could be also the (2, $-$2) mode, but this case is quite tricky.
In the lower limit of the rotation rate, the (2, $-$2) mode is unstable at the frequency $\nu_1$.
When the rotation increases, the instability domain shifts towards lower frequencies and the mode (2, $-$2) becomes stable.
Finally, for sufficiently high values of $V_\mathrm{rot}$, the mode (2, $-$2) is reflected and
its instability domain again encompasses the frequency $\nu_1$.
The frequency $\nu_2$ is close to $\nu_1$ and therefore its instability
is approximately the same.
The third frequency, $\nu_3$, is the highest. It can be associated with
the modes (1, 0), (2, $-$2) for the highest rotation rates, with (1, +1), (2, $-$1), (2, 0), (r, $-$2)
for the slow and moderate rotation rates, and with (2, +1) for the slowest rotation.
The lowest observed frequency is $\nu_4$. In the regime of the low
rotation velocity, it can be the (1, $-$1) or (1, +1) mode.
In the case of the low to moderate rotation rate it can be the dipole axisymmetric
or the (r, $-$2) mode. $\nu_4$  can also be the (2, $-$2) mode for the lowest and high rotation rates
and the (2, $-$1) mode in the whole range of the $V_\mathrm{rot}$.

The (2, +2) mode cannot account for any of the observed frequencies
because its instability occurs at higher frequencies than the observed ones.
None of the peaks may be associated with the (r, $-$1) mode.
It is worth to emphasize that r modes are unstable only in narrow ranges of frequencies.

In the next step, we computed the values of $\chi^2$ using equation\,(4).
As an example, we present results for the dominant frequency (see Fig.\,\ref{chi_rot}, {top panel}).
Five unstable modes met our $\chi^2<1.5$ condition.
The angular numbers ($\ell,~m$) of these modes and the ranges of $V_\mathrm{rot}$
are listed in the third column of Table\,\ref{mic_tab_HD21071}. We called this approach the \emph{a} case.

In the \emph{b} case, we rejected modes with the empirical values of $|\varepsilon|$
above the adopted maximum of 0.01.
As one can see from Fig.\,\ref{chi_rot} {(the top and middle panels)},
in the ranges of the rotational velocity in which the value of $\chi^2$ is acceptable,
the modes (2,0), (3, $-$1) and (4, $-$2) have too large values of $|\varepsilon|$.
The modes with the appropriate values of $\chi^2$ and $|\varepsilon|$ are (1, 0), (1, +1);
they are listed in the fourth column of Table\,\ref{mic_tab_HD21071}.

Another constraint used by us is the value of the amplitude of the radial velocity
changes, $A_{V_{\rm rad}}$, and this is the \emph{c} case. In the case
of HD\,21071, only the dominant frequency was detected in
the spectroscopic data. In Fig.\,\ref{chi_rot} {(the bottom panel)} are shown the theoretical values of $A_{V_{\rm rad}}$
calculated with the values of $\varepsilon$ obtained from equation\,(3) for modes presented in {the top panel}.
As one can see, the (1, +1) mode should be rejected due to the excessive value of the $A_{V_{\rm rad}}$,
while the (1, 0) mode has the theoretical value of $A_{V_{\rm rad}}$ within the observational error
for $V_\mathrm{{rot}}$ above 211 km s$^{-1}$. In the case of the other frequencies,
we required that their theoretical values of $A_{V_{\rm rad}}$ were smaller than the observed $A_{V_{\rm rad}}$
for the dominant frequency.
These identifications are given in the fifth column of Table\,\ref{mic_tab_HD21071}.

Finally, we selected only modes with a common range of $V_\mathrm{rot}$ for all frequencies -- the \emph{d} case.
This gives us a very narrow range of the rotational velocities,
$V_\mathrm{{rot}}\in [211;212]$ km s$^{-1}$ and unambiguous mode identification for all observed
frequencies: $\nu_1$, $\nu_2$, $\nu_3$ are dipole axisymmetric modes and
$\nu_4$ is a quadrupole mode with $m=-1$ (the last column in Table \,\ref{mic_tab_HD21071}).

To check the dependence of our mode identification on the adopted model parameters,
the above described procedure was repeated for the models corresponding to the
four edges of the error box {and for the model from the centre of the error box
calculated with overshooting from the convective core taken into account, $\alpha_\mathrm{ov}=0.2\,H_p$}.
The results are presented in Table\,\ref{mie_tab_HD21071}.
Depending on the model parameters, we obtained the rotation velocity from
about 150  to 250 km s$^{-1}$.

For Model\#1 and Model\#2, we obtained additional solutions for $\nu_4$ and the smaller values of $V_\mathrm{{rot}}$.
In the case of Model\#3, we were unable to find mode identification for $\nu_4$ in the common range of $V_\mathrm{rot}$
with other frequencies. Therefore, in Table\,\ref{mie_tab_HD21071} are listed in italics
the three modes with the values of $V_\mathrm{{rot}}$ closest to the range of $V_\mathrm{{rot}}$
obtained for $\nu_1$, $\nu_2$ and $\nu_3$.
The most complicated situation is for the coolest and most luminous Model\#4.
In this model, for $\nu_1$  we were able to find the modes which fit the amplitude of the radial velocity
changes only to within 2$\sigma$ (in Table\,\ref{mie_tab_HD21071} indicated as $2\sigma$)
and it gave us two solutions. Another problem is the fact that in the case of the solution for $\nu_1$, which is consistent
with other models, namely the (1, 0) mode, there is no common range of $V_\mathrm{{rot}}$ with the results for $\nu_4$
(the three modes with the closest rotational velocity ranges are given in Table\,\ref{mie_tab_HD21071} in italic).
{In the case of the model with core overshooting (Model\#5), as for Model\#3,
we were unable to find mode identification for $\nu_4$ in the common range of $V_\mathrm{rot}$ with other frequencies.

It is worth to mention that in more evolved stars overshooting from the convective core may
affect a model more significantly than in the case of HD\,21071.
Moreover, there is a well-known problem in the literature that discrepancies
between different determinations of the fundamental parameters exist \citep[e.\,g.][]{Balona2011}.
These uncertainties can affect mode identification.
From Table\,\ref{mie_tab_HD21071}, one can see that the results of our mode identification, especially
the values of the rotational velocity, depend on the adopted model.
Uncertainties in our results are discussed in more details in the last section.}

\begin{table*}
 \centering
 \begin{minipage}{\textwidth}
  \caption{The results of the mode identification of the frequency peaks observed
              in HD\,21071 based on the evolutionary models from the four
              edges of the observational error box and from the centre
              of the observational error box but with overshooting from
              convective core taken into account.}
\label{mie_tab_HD21071}
\centering
  \begin{tabular}{l|ll|ll|ll|ll|ll}
  \hline
        &  \multicolumn{10}{|c}{$\ell$($V_\mathrm{rot}$=0)~~~~~~$\left(\ell,m\right) \left[V_\mathrm{rot}^\mathrm{min};V_\mathrm{rot}^\mathrm{max}\right]$ km s$^{-1}$} \\
      &  \multicolumn{10}{|c}{}\\
\cline{2-11} \\
$\nu$ (d$^{-1}$)      & \multicolumn{2}{|c|}{Model\#1} & \multicolumn{2}{|c|}{Model\#2} & \multicolumn{2}{|c|}{Model\#3} & \multicolumn{2}{|c}{Model\#4}& \multicolumn{2}{|c}{Model\#5}\\

 \hline

1.18843 & 1  & (1,0) [148;170] & 1  & (1,0) [174;200] & 1 & (1,0) [249;255]       & 1 & (1,0) [229,245]$^{2\sigma}$ & 1  & (1,0) [212;235] \\
        &    &                 &    &                 &   &                       &   & (1,1) [107;110]$^{2\sigma}$ &    & \\
\hline
1.14934 & 1  & (1,0) [148;170] & 1  & (1,0) [174;200] & 1 & (1,0) [249;255]       & 1 & (1,0) [229,245]             & 1  & (1,0) [212;235]\\
        & 2  &                 & 2  &                 &   &                       &   & (1,1) [107;110]             &    &\\
\hline
1.41968 & 1  & (1,0) [148;170] & 1  & (1,0) [174;200] & 1 & (1,0) [249;255]       & 1 & (1,0) [229,245]             & 1  & (1,0) [212;235]\\
        & 2  &                 &    &                 &   &                       &   & (1,1) [107;110]             &    &\\
        &    &                 &    &                 &   &                       &   & (2,0) [108;110]             &    &\\
\hline
0.95706 & 1  & (1,-1) [148;170]& 1  & (1,-1) [174;200]& 1 &\emph{(1,0) [50;180]}  & 1 & \emph{(1,0) [50;215]}       &1   & \emph{(1,0) [50;155]}\\
        & 2  & (2,-1) [148;170]& 2  & (2,-1) [174;200]& 2 &\emph{(2,-1) [50;189]} & 2 & \emph{(2,-1) [50;177]}      & 2  & \emph{(2,-1) [50;210]}\\
        &    & (2,-2) [148;170]&    & (2,-2) [180;190]&   &\emph{(2,-2) [180;189]}&   & \emph{(2,-2) [185;192]}     &    & \emph{(2,-2) [175;188]}\\
        &    &                 &    &                 &   &                       &   & (1,0) [107;110]             &    &\\
        &    &                 &    &                 &   &                       &   & (2,-1) [107;110]            &    &\\

\hline
\end{tabular}
\footnotetext{
$^{2\sigma}$ -- fit A$_{V_\mathrm{rad}}$ within 2 $\sigma$; {in italics are printed
modes without a common
range of $V_\mathrm{rot}$ with other frequency peaks}\\
Model\#1 -- $\log T_\mathrm{eff}$=4.171; $\log L/\mathrm L_\odot$=2.368; $M=3.62\mathrm M_\odot$; $\alpha_\mathrm{ov}=0.0\,H_p$;\\
Model\#2 -- $\log T_\mathrm{eff}$=4.157; $\log L/\mathrm L_\odot$=2.368; $M=3.55\mathrm M_\odot$; $\alpha_\mathrm{ov}=0.0\,H_p$;\\
Model\#3 -- $\log T_\mathrm{eff}$=4.171; $\log L/\mathrm L_\odot$=2.520; $M=3.84\mathrm M_\odot$; $\alpha_\mathrm{ov}=0.0\,H_p$;\\
Model\#4 -- $\log T_\mathrm{eff}$=4.157; $\log L/\mathrm L_\odot$=2.520; $M=3.78\mathrm M_\odot$; $\alpha_\mathrm{ov}=0.0\,H_p$;\\
Model\#5 -- $\log T_\mathrm{eff}$=4.164; $\log L/\mathrm L_\odot$=2.444; $M=3.64\mathrm M_\odot$; $\alpha_\mathrm{ov}=0.2\,H_p$;\\
}
\end{minipage}
\end{table*}

\subsection{Other results}

\begin{table*}
 \centering
 \begin{minipage}{\textwidth}
  \caption{The results of the mode identification for the frequency peaks in all programme stars.
               The values of the rotational velocity and frequency are in km s$^{-1}$ and d$^{-1}$, respectively.
              In the last column (in the section `stars with radial velocity variations'),
              the spectroscopically determined values of $\left(\ell\,,m\right)$
              are given if available \citep{DeCat2005}.
               }
\label{tab_results_all}
\centering
  \begin{tabular}{llllll}
  \hline

HD      & $\nu$ (d$^{-1}$) & $\ell$($V_\mathrm{rot}$=0) &
\multicolumn{2}{c}{$\left(\ell,m\right) \left[V_\mathrm{rot}^\mathrm{min};V_\mathrm{rot}^\mathrm{max}\right]  \left\{\nu_\mathrm{rot}^\mathrm{min};\mathrm \nu_\mathrm{rot}^\mathrm{max} \right\}$} & $\left(\ell,m\right)$\\

 \hline

 \multicolumn{6}{c}{Stars with radial velocity variations} \\
 \hline

24587    & 1.1569   & 1            &  (1,1) [115;127] \{0.6952;0.7677\}                                &    & (1,1)\\
\hline
25558    & 0.65265  & 1            & (1,0) [174;198] \{0.8396;0.9554\}$^{2\sigma}$                     &   \\
         & 1.93235  &  ?           & (6,-1) [174;198] \{0.8396;0.9554\}                                &   \\
         & 1.17913  & 1, 2, 3      & (4,-1) [174;198] \{0.8396;0.9554\}                                & (6,-1) [174;198] \{0.8396;0.9554\}  \\
\hline
26326    & 0.5338   &1, 2          & (1,0) [71;92]  \{0.3653;0.4733\}                                  &  \\
         & 0.7629   & 1            & (1,1) [71;92]  \{0.3653;0.4733\}                                  &  \\
\hline
53921    & 0.6054   & 1!           &  (2,-1) [17;77] \{0.1209;0.5475\}!                                &   \\
\hline
74195    & 0.35475  &1, 2          & (1,0) [25;43] \{0.0986;0.1696\}$^{6\sigma}$                       & (1,-1) [25;29] \{0.0986;0.1144\}$^{6\sigma}$ \\
         & 0.35033  &1, 2          & (1,0) [25;43] \{0.0986;0.1696\}$^{2\sigma}$                       & \\
         & 0.34630  &1, 2          &  (2,-2) [25;43] \{0.0986;0.1696\}                                 &    \\
         & 0.39864  &1, 2          & (2,-1) [40;43] \{0.1578;0.1696\}                                  & (2,-2) [25;40] \{0.0986;0.1578\}  \\
\hline
74560    & 0.6447137& 1            &  (1,0) [79;135] \{0.4621;0.7897\}                                 & (1,-1) [60;73] \{0.3510;0.4270\} \\
         & 0.3957710& 1            &  (1,-1) [60;73] \{0.3510;0.4270\}                                 & (1,-1) [79;135] \{0.4621;0.7897\} \\
         &          &              &  (2,-1) [115;135] \{0.6727;0.7897\}                               & (2,-2) [60;73] \{0.3510;0.4270\} \\
         &          &              &  (2,-2) [110;133] \{0.6435;0.7780\}   \\
         & 0.4476463&1             &  (1,-1) [60;73]  \{0.3510;0.4270\}                                & (1,-1) [79;135]  \{0.4621;0.7897\}  \\
         &          &              &  (2,-1) [60;73]  \{0.3510;0.4270\}                                & (2,-1) [79;135]  \{0.4621;0.7897\} \\
         &          &              &  (2,-2) [60;73]   \{0.3510;0.4270\}                               & (2,-2) [79;90]   \{0.4621;0.5265\} \\
         &          &              &  (2,-2) [120;135] \{0.7020;0.7897\}                               & (4,-3) [90;91]  \{0.5265;0.5323\}         \\
         & 0.6359820&1, 2          &   ?                                                               &  \\
         & 0.8228450& 1            &   ?                                                               &  \\
         & 1.6105673&1, 2          &   ?                                                               &  \\
\hline
85953    & 0.2663   & 1            &  (1,-1) [169;171] \{0.5085;0.5145\}                               &  \\
         & 0.2189   &1             & (1,-1) [169;171] \{0.5085;0.5145\}                                & (6,-2) [169;171] \{0.5085;0.5145\} \\
\hline

92287     & 0.21480  & 1            & (1,-1) [122-134]  \{0.5002;0.5494\}                              &  & (1,-1)\\
\hline
121190    & 2.6831   &  ?           & (1,1) [198-233] \{1.7238;2.0285\}                                & \\
          & 2.6199   &  ?           & ?                                                                & \\
          & 2.4713   &  ?           & (1,1) [198-233] \{1.7238;2.0285\}                                & \\
\hline
123515    & 0.68528  & 1            &  \emph{(1,0) [117;149]} \{0.5757;0.7332\}                        & \\
          & 0.65929  & 1            & (1,0) [166;194]  \{0.8169;0.9546\}                               & \\
          & 0.72585  & 1            &  (1,0) [166;194] \{0.8169;0.9546\}                               &  \\
          & 0.55198  &1, 4          &  \emph{(1,0) [24;149]}    \{0.1181;0.7332\}                      &  \\
\hline
138764    & 0.7944   & 1, 4         & (1,0) [141;163] \{0.9365;1.0826\}                                & (1,1) [53;64]   \{0.3520;0.4251\} \\
\hline
140873    & 1.1515   & 1            &  (3,0) [101;102] \{0.5921;0.5980\}                               & (4,0) [80;81]   \{0.4690;0.4749\} & (1,1)\\
\hline
177863    & 0.84059  & 1            &  (2,0)  [62;63] \{0.4074;0.4140\}                                & (2,0) [68;69] \{0.4469;0.4534\} & (1,1)\\
          & 0.10108  & ?            &  (2,-2) [62;63] \{0.4074;0.4140\}                                & (2,-2) [68;69] \{0.4469;0.4534\} \\
          &          &              &  (4,-3) [62;63]  \{0.4074;0.4140\}                               &    \\
\hline
181558    & 0.80780  & 1            & (1,0) [228;242] \{1.2699;1.3479\}                                & (1,1) [77;80]  \{0.4289;0.4456\} & (1,1)\\
\hline
182255    & 0.97185  & 1            & (1,1) [54;75] \{0.3921;0.5446\}$^{6\sigma}$                      & \\
          & 0.79225  &1             &   (1,0) [54;75] \{0.3921;0.5446\}                                &  \\
          & 0.62526  & 1, 2         &  (1,0) [54;65] \{0.3921;0.4720\}                                 & (1,-1) [54;75] \{0.3921;0.5446\} \\
          &          &              &  (2,-1) [54;75] \{0.3921;0.5446\}                                & \\
          & 1.12780  &  1           &  (1,0) [54;75] \{0.3921;0.5446\}                                 & (1,1) [54;75] \{0.3921;0.5446\} \\
          &          &              &  (2,0) [54;75] \{0.3921;0.5446\}                                 &  \\
          & 1.02884  & 1, 2         &   (1,0) [54;75] \{0.3921;0.5446\}                                & (1,1) [54;75] \{0.3921;0.5446\} \\
          &          & 4            &   (2,0) [54;75] \{0.3921;0.5446\}                                &     \\
\hline
208057    & 0.89050  & 1            & \emph{(2,0) [120-122]} \{0.5935;0.6034\}                         &  \\
          & 0.80213  &  1, 2        &  (1,0) [138;177] \{0.6825;0.8754\}                               & \\
          & 2.47585  &  ?           & (4,0) [138;177] \{0.6825;0.8754\}                                & \\
\hline
215573    & 0.5654   & 1, 2         & (1,0) [75;93]  \{0.4535;0.5623\}                                 & (1,1) [28;33]  \{0.1693;0.1995\}\\
\hline

\end{tabular}
\end{minipage}
\end{table*}


\begin{table*}
 \centering
 \begin{minipage}{\textwidth}
  \contcaption{}
\centering
  \begin{tabular}{lllll}
  \hline

HD      & $\nu$ (d$^{-1}$) & $\ell$($V_\mathrm{rot}$=0) &
\multicolumn{2}{c}{$\left(\ell,m\right) \left[V_\mathrm{rot}^\mathrm{min};V_\mathrm{rot}^\mathrm{max}\right]  \left\{\nu_\mathrm{rot}^\mathrm{min};\mathrm \nu_\mathrm{rot}^\mathrm{max} \right\}$} \\

 \hline
\multicolumn{5}{c}{Stars without radial velocitiy variations}	\\
\hline

1976     & 0.39946  & 1            &  $\ell \le$5 [143;223] \{0.4770;0.7438\}                          &  \\
         & 0.93895  &      1       &  (1,1) [163;223] \{0.5437;0.7438\}                                & (2,1) [144;163] \{0.4803;0.5437\}  \\
         &          &              &  (4,0) [143;152] \{0.4770;0.5070\}                                & (6,0) [143;146] \{0.4770;0.4870\}  \\
         & 1.20346  & 1, 2, 3      &   $\ell \ne$1 [143;223] \{0.4770;0.7438\}                         &  \\
\hline
3379     & 1.82023  & 1, 2         &  $\ell \ne$1                                                      &   \\
         &1.59418   & 1            &  $\ell \ne$1                                                      &    \\
\hline
28114    & 0.48790  & 1            & (1,0) [21;111] \{0.1123;0.5935\}                                  & (1,-1) [21;45] \{0.1123;0.2406\}   \\
         &          &              & (1,1)  [21;41] \{0.1123;0.2192\}                                  & (2,0) [21;25]  \{0.1123;0.1337\} \\
         &0.49666  &1              &  (1,0) [21;111] \{0.1123;0.5935\}                                 & (1,-1) [21;92] \{0.1123;0.4919\} \\
         &         &               &  (1,1) [21;41]  \{0.1123;0.2192\}                                 & (2,0) [21;25]  \{0.1123;0.1337\} \\
         &         &               &  (2,-1) [21;60]  \{0.1123;0.3208\}                                & (3,-1) [73;101] \{0.3903;0.5400\} \\
\hline
28475    & 0.68369  &1, 2          & (1,0) [30;165] \{0.1810;0.9956\}                                  & (1,-1) [30;74] \{0.1810;0.4465\} \\
         &          &              & (1,1) [30;55]  \{0.1810;0.3319\}                                  & (2,-1) [30;63] \{0.1810;0.3802\} \\
         &          &              & (2,0) [30;36] \{0.1810;0.2172\}                                   & (2,0) [38;50] \{0.2293;0.3017\} \\
         &          &              & (3,-1) [41;42] \{0.2474;0.2534\}                                  & (4,-1) [85;206] \{0.5129;1.2430\} \\
         &  0.40893 &1             & (1,-1) [30;203] \{0.1810;1.2249\}                                 & (2,-1) [105;156] \{0.6336;0.9413\} \\
         &          &              & (2,-2) [30;56] \{0.1810;0.3379\}                                  & (2,-2) [110;116] \{0.6638;0.7000\}  \\
         &          &              & (4,-2) [50;54] \{0.3017;0.3258\}                                  & (4,-3) [31;36] \{0.1871;0.2172\} \\
         &          &              & (r,-1) [165;206] \{0.9956;1.2430\}                                &  \\
\hline

34798    & 0.78350  & 1            & (1,0) [31;157]  \{0.1878;0.9513\}                                 & (1,-1) [30;49]  \{0.1818;0.2969\} \\
         &          &              & (1,1) [30;65]   \{0.1818;0.3939\}                                 & (2,0) [30;34] \{0.1818;0.2060\}  \\
         &          &              & (2,0) [40;70] \{0.2424;0.4242\}                                   & \\
         & 0.85231  &1, 2, 4       & ?                                                                 & \\
         &0.97432   & 1            & (1,0) [31;157] \{0.1878;0.9513\}                                  & (1,1) [30;100] \{0.1818;0.6059\} \\
         &          &              & (2,0) [30;35]  \{0.1818;0.2121\}                                  & (2,0) [39;120]  \{0.2363;0.7271\} \\
         &          &              & (2,1) [30;50]  \{0.1818;0.3030\}                                  & (2,2) [30;30]  \{0.1818;0.1818\} \\
         &          &              & (3,0) [44;70]  \{0.2666;0.4242\}                                  & (4,-1) [89;157]  \{0.5393;0.9513\} \\
         &          &              & (5,-1) [129;157] \{0.7817;0.9513\}                                & (6,-6) [30;30]  \{0.1818;0.1818\} \\
         & 0.67795  &1, 2          & (1,0) [30;145]  \{0.1818;0.8786\}                                 & (1,-1) [30;101] \{0.1818;0.6120\} \\
         &          &              & (1,1) [30;45]   \{0.1818;0.2727\}                                 & (2,0) [30;35]   \{0.1818;0.2121\} \\
         &          &              & (2,-1) [30;75]  \{0.1818;0.4545\}                                 & (3,-1) [36;157] \{0.2181;0.9513\} \\
         &          &              & (4,-2)[30;30]   \{0.1818;0.1818\}                                 & \\
\hline
37151    & 1.2431665& 1!           & (1,0) [20;100]  \{0.2195;1.0974\}                                 & (1,-1) [20;140] \{0.2195;1.5363\} \\
         &          &              & (1,1) [20;45]   \{0.2195;0.4938\}                                 & \\
         & 1.1801861& 1            & (1,0) [20;85]   \{0.2195;0.9328\}                                 & (1,-1) [20;140] \{0.2195;1.5363\} \\
         &          &              & (1,1) [20;40]   \{0.2195;0.4390\}                                 & \\
         & 1.0426632& 1            & (1,0) [20;60]   \{0.2195;0.6584\}                                 & (1,-1) [20;140] \{0.2195;1.5363\} \\
         &          &              & (1,1) [20;25]   \{0.2195;0.2743\}                                 & \\
         & 1.1052525&  1           & (1,0) [20;70]   \{0.2195;0.7682\}                                 & (1,-1) [20;140] \{0.2195;1.5363\} \\
         &          &              & (1,1) [20;30]   \{0.2195;0.3292\}                                 & (2,-2) [30;35]  \{0.3292;0.3841\} \\
         & 1.4520631&  1, 2        & (1,0) [20;140]  \{0.2195;1.5363\}                                 & (1,-1) [20;70]  \{0.2195;0.7682\} \\
         &          &              & (1,1) [20;65]   \{0.2195;0.7133\}                                 & (2,-1) [20;105] \{0.2195;1.1522\} \\
         &          &              & (2,-2) [20;68]  \{0.2195;0.7462\}                                 & \\


\hline

45284    & 1.23852  & 1            & (1,0) [53;272]    \{0.4083;2.0954\}                               & (1,1) [52;102]    \{0.4006;0.7858\} \\
         &          &              & (2,0) [52;60] \{0.4006;0.4622\}                                   & (2,0) [67;117] \{0.5161;0.9013\} \\
         &          &              & (3,0) [62;67]     \{0.4776;0.5161\}                               & (4,-1) [103;272]  \{0.7935;2.0954\} \\
         &          &              & (5,-1) [119;152]  \{0.9167;1.1709\}                               &  \\
         & 1.12753  & 1            & (1,0) [53;272]    \{0.4083;2.0954\}                               & (1,-1) [52;52]    \{0.4006;0.4006\}\\
         &          &              & (1,1) [52;87]     \{0.4006;0.6702\}                               & (2,0) [52;60]  \{0.4006;0.4622\} \\
         &          &              & (2,0) [68;97]  \{0.5238;0.7472\}                                  & (4 ,-1) [106;192] \{0.8166;1.4791\} \\
         & 1.50605  & ?            & ?                                                                 &           \\
         &              &              &                                                                    &          \\
\hline

\end{tabular}
\end{minipage}
\end{table*}


\begin{table*}
 \centering
 \begin{minipage}{\textwidth}
  \contcaption{}
\centering
  \begin{tabular}{lllll}
  \hline

HD      & $\nu$ (d$^{-1}$) & $\ell$($V_\mathrm{rot}$=0) &
\multicolumn{2}{c}{$\left(\ell,m\right) \left[V_\mathrm{rot}^\mathrm{min};V_\mathrm{rot}^\mathrm{max}\right]  \left\{\nu_\mathrm{rot}^\mathrm{min};\mathrm \nu_\mathrm{rot}^\mathrm{max} \right\}$} \\

\hline
143309   & 0.59830  & 1           & (1,0) [10;121]   \{0.0611;0.7392\}                                 & (1,-1) [10;19]   \{0.0611;0.1161\} \\
         &          &             & (1,1) [10;45]    \{0.0611;0.2749\}                                 & (2,-1) [10;30]   \{0.0611;0.1833\} \\
         &          &             & (2,0) [10;11]  \{0.0611;0.0672\}                                   & (2,0) [14;30]  \{0.0855;0.1833\} \\
         &          &             & (2,-2) [10;12]   \{0.0611;0.0733\}                                 & \\
         & 0.60213  & 1, 2        & (1,0) [10;121]   \{0.0611;0.7392\}                                 & (1,-1) [10;22]   \{0.0611;0.1344\} \\
         &          &             & (1,1) [10;45]    \{0.0611;0.2749\}                                 & (2,0) [10;11]  \{0.0611;0.0672\}\\
         &          &             & (2,0) [13;30]  \{0.0794;0.1833\}                                   & (2,-1) [10;40]   \{0.0611;0.2444\} \\
         &          &             & (2,-2) [10;14]   \{0.0611;0.0855\}                                 & \\
         & 0.71414  & 1, 2        & (1,0) [10;121]   \{0.0611;0.7392\}                                 & (1,-1) [10;23]   \{0.0611;0.1405\}\\
         &          &             & (1,1) [10;65]    \{0.0611;0.3971\}                                 & (2,0) [10;11]  \{0.0611;0.0672\} \\
         &          &             & (2,0) [13;60]  \{0.0794;0.3666\}                                   & (2,-1) [10;42]   \{0.0611;0.2566\} \\
         &          &             & (2,1) [10;25]    \{0.0611;0.1527\}                                 & (2,-2) [10;15]   \{0.0611;0.0916\}\\
         &          &             & (2,2) [10;15]    \{0.0611;0.0916\}                                 & \\
         & 1.20066  & 1           & (2,0) [74;121]   \{0.4521;0.7392\}                                 & (2,1) [29;90]    \{0.1772;0.5498\}\\
         &          &             & (3,0) [63;115]   \{0.3849;0.7026\}                                 & (4,0)[71;80]     \{0.4338;0.4887\} \\
         &          &             & (4,1) [21;35]    \{0.1283;0.2138\}                                 & (4,2) [14;20]    \{0.0855;0.1222\} \\
         &          &             & (4,4) [10;10]    \{0.0611;0.0611\}                                 & \\
         & 0.59133  & 1, 2        & (1,0) [10;120]   \{0.0611;0.7331\}                                 & (1,-1) [10;54]   \{0.0611;0.3299\} \\
         &          &             & (1,1) [10;45]    \{0.0611;0.2749\}                                 & (2,0) [10;12]  \{0.0611;0.0733\} \\
         &          &             & (2,0) [13;30] \{0.0794;0.1833\}                                    & (2,-1) [10;121]  \{0.0611;0.7392\} \\
         &          &             & (2,-2) [10;22]   \{0.0611;0.1344\}                                 & (3,-1) [35;95]   \{0.2138;0.5804\} \\

\hline
160124   & 0.52014  & 1           & (1,0) [11;78]    \{0.0439;0.3113\}           &  (1,1) [8;11]  \{0.0319;0.0439\} \\
         &          &             & (1,1) [74;78]  \{0.2954;0.3113\}             &  (2,0) [8;8]   \{0.0319;0.0319\} \\
         &          &             & (2,0) [13;78]  \{0.0519;0.3113\}             &  (2,1) [8;38]     \{0.0319;0.1517\} \\
         &          &             & (2,2) [8;14]     \{0.0319;0.0559\}           &  (3,0) [40;58]    \{0.1596;0.2315\} \\
         & 0.52096  & 1           & (1,0) [10;78]    \{0.0399;0.3113\}           &  (1,1) [8;14]  \{0.0319;0.0559\} \\
         &          &             & (1,1) [63;78]  \{0.2514;0.3113\}             &  (2,0) [8;8]   \{0.0319;0.0319\} \\
         &          &             & (2,0) [12;78]   \{0.0479;0.3113\}            &  (2,1) [8;38]     \{0.0319;0.1517\} \\
         &          &             & (2,2) [8;17]     \{0.0319;0.0679\}           &  (3,0) [36;58]    \{0.1437;0.2315\} \\
         & 0.52055  & 1           & (1,0) [12;78]    \{0.0479;0.3113\}           &  (1,1) [8;10] \{0.0319;0.0399\} \\
         &          &             & (1,1) [76;78] \{0.3033;0.3113\}              &  (2,0) [13;78]    \{0.0519;0.3113\}\\
         &          &             & (2,1) [8;38]     \{0.0319;0.1517\}           &  (2,2) [8;14]     \{0.0319;0.0559\} \\
         &          &             & (3,0) [42;58]    \{0.1676;0.2315\}           &   \\
         & 0.52137  & ?           &  ?                                           &   \\
         & 0.52197  & 1           &  (1,0) [11;78]   \{0.0439;0.3113\}           & (1,1) [8;11] \{0.0319;0.0439\} \\
         &          &             &  (1,1) [74;78] \{0.2954;0.3113\}             & (2,0) [8;8]  \{0.0319;0.0319\} \\
         &          &             &  (2,0) [13;78]  \{0.0519;0.3113\}            & (2,1) [8;38]    \{0.0319;0.1517\} \\
         &          &             &  (2,2) [8;14]    \{0.0319;0.0559\}           & (3,0) [42;58]   \{0.1676;0.2315\} \\
         & 0.70259  &  1          & (2,0) [8;8]   \{0.0319;0.0319\}              & (2,0) [13;30]  \{0.0519;0.1197\}      \\
         &          &             & (2,-1) [9;26]    \{0.0359;0.1038\}           & (2,1) [59;78]    \{0.2355;0.3113\}  \\
         &          &             & (2,-2) [8;10]    \{0.0319;0.0399\}           & (4,0) [64;78]    \{0.2554;0.3113\} \\
         & 1.0422   & 1, 2, 3, 4  & $\ell \ne$1                                  &      \\
         & 0.31366  &  1, 2       & (1,0) [8;78]     \{0.0319;0.3113\}           & (1,-1) [8;45]    \{0.0319;0.1796\}  \\
         &          &             & (1,1) [8;23]     \{0.0319;0.0918\}           & (2,-1) [8;67]    \{0.0319;0.2674\}  \\
         &          &             & (2,-2) [8;19]    \{0.0319;0.0758\}           &   \\
\hline
  160762 & 0.28671  & ?           & (1,-1) [98;105]  \{0.4414;0.4729\}           & \\
\hline
179588   & 0.85654  & 1, 4        &  \emph{(1,0) [125;205] \{0.6481;1.0628\}}    & \\
         & 2.04263  & 1!, 2!      &  \emph{(2,1) [177;190] \{0.9177;0.9851\}!!}  & \\
         & 2.19989  & 1           &  \emph{(2,1) [203;205] \{1.0525;1.0628\}}    & \\
         & 1.83359  & 1           &  \emph{(2,1) [135;165] \{0.6999;0.8555\}}    &  \\
\hline
191295   & 0.71908  & 1           & (1,0) [17;128]   \{0.1033;0.7774\}           & (1,-1) [16;32]   \{0.0972;0.1944\} \\
         &          &             & (1,1) [16;71]    \{0.0972;0.4312\}           & (2,0) [37;56]    \{0.2247;0.3401\} \\
         &          &             & (2,1) [16;26]    \{0.0972;0.1579\}           & \\
         & 0.49011  & 1, 2        &  (1,0) [16;76]   \{0.0972;0.4616\}           & (1,-1) [16;128] \{0.0972;0.7774\} \\
         &          &             &  (1,1) [16;31]   \{0.0972;0.1883\}           & (2,-1) [16;81]  \{0.0972;0.4920\} \\
         &          &             &  (2,-2) [16;35]  \{0.0972;0.2126\}           &  \\
         & 0.29302  &1            &  (1,-1) [21;128] \{0.1275;0.7774\}           & (2,-2) [31;49]  \{0.1883;0.2976\} \\
\hline

\end{tabular}
\end{minipage}
\end{table*}


\begin{table*}
 \centering
 \begin{minipage}{\textwidth}
  \contcaption{}
\centering
  \begin{tabular}{lllll}
  \hline

HD      & $\nu$ (d$^{-1}$) & $\ell$($V_\mathrm{rot}$=0) &
\multicolumn{2}{c}{$\left(\ell,m\right) \left[V_\mathrm{rot}^\mathrm{min};V_\mathrm{rot}^\mathrm{max}\right]  \left\{\nu_\mathrm{rot}^\mathrm{min};\mathrm \nu_\mathrm{rot}^\mathrm{max} \right\}$} \\

\hline

206540   & 0.72002  & 1           &  (1,0) [17;149]  \{0.0932;0.8165\}           & (1,0) [160;177]  \{0.8768;0.9699\}  \\
         &          &             &  (1,-1) [15;21]  \{0.0822;0.1151\}           & (1,1) [15;85]   \{0.0822;0.4658\} \\
         &          &             &  (2,0) [59;85]   \{0.3233;0.4658\}           & (2,1) [21;40]   \{0.1151;0.2192\} \\
         &          &             &  (2,2) [15;25]   \{0.0822;0.1370\}           & \\
         & 0.62125  & 1, 2, 4     &  (1,0) [15;149]   \{0.0822;;0.8165\}         & (1,0) [160;177] \{0.8768;0.9699\} \\
         &          &             &  (1,-1) [15;35]  \{0.0822;0.1918\}           & (1,1) [15;65]   \{0.0822;0.3562\} \\
         &          &             &  (2,0) [15;17]       \{0.0822;0.0932\}       &  (2,0) [20;60]  \{0.1096;0.3288\}    \\
         &          &             &  (2,-1) [15;76]  \{0.0822;0.4165\}           & (2,1) [15;25]   \{0.0822;0.1370\} \\
         &          &             &  (2,-2) [15;21]  \{0.0822;0.1151\}           & (2,2) [15;15]   \{0.0822;0.0822\} \\
         & 0.38271  &1, 2         &  (1,0) [15;55]   \{0.0822;0.3014\}           & (1,-1) [15;149] \{0.0822;0.8165\} \\
         &          &             &  (1,1) [15;15]   \{0.0822;0.0822\}           & (2,-1) [25;75]  \{0.1370;0.4110\} \\
         &          &             &  (2,-2) [15;36]  \{0.0822;0.1973\}           & (r,-1) [160;177] \{0.8768;0.9699\} \\


\hline
\end{tabular}
\footnotetext{
$^{2\sigma}$ -- radial velocity within  2$\sigma$; $^{6\sigma}$ -- radial velocity within 6$\sigma$;\\
! -- accepted  $\chi^2 \le 3.3$; !! -- accepted  $\chi^2 \le 4.6$;\\
? -- no constraints;
{in italics are printed modes without a common range of $V_\mathrm{rot}$ with other frequency peaks.}
}
\end{minipage}
\end{table*}

The results of mode identification for the remaining programme stars are presented in Table\,\ref{tab_results_all}.
It is clearly seen that the best constraints on ($\ell,~m$) and the rotational velocity
were obtained for the stars with the measured radial velocity variations.

Most of the stars we studied in this paper pulsate in the normal modes. There are two exceptions:
$\nu_2$ in HD\,28475 and $\nu_3$ in HD\,206540 which may be either an r mode or a normal g mode.

In the case of $\nu_1$ in HD\,53921 (identification with and without rotation), $\nu_1$ in HD\,37151
(identification without rotation) and $\nu_2$ in HD\,179588 (identification with and without rotation),
we are not able to find solutions with $\chi^2\le$1.5 and therefore we accepted higher values of
$\chi^2$ (indicated in Table \ref{tab_results_all} with an exclamation mark
for $\chi^2_\mathrm{max}$=3.3 which corresponds to the probability
that real solution has higher $\chi^2$ equal to 1 per cent or by a double exclamation mark for $\chi^2_\mathrm{max}$=4.6
which corresponds to the probability that real solution has higher $\chi^2$ equal to 0.1 per cent).
We would like to remind that these quantities should be treated more like
arbitrarily chosen thresholds than a strict statistical approach.

In the case of spectroscopic binaries, it was checked
whether the pulsation frequency can be a multiple of the orbital frequency.
Only for the frequency $\nu_1$ in the star HD\,177863,
the ratio of the pulsation frequency to the orbital frequency,
equal to 10.016$\pm$0.001, is close, but still not equal to, within observational error, the integer.
A possibility that this mode is excited by tidal forces was studied
by \citet{Willems_Aerts2002} but they did not conclude
which mechanism, classical $\kappa$ or tidal forces,
is responsible for driving this mode.
Another possibility is that some observed frequencies are the result of rotational modulation
due chemical and/or temperature inhomogeneities. The frequencies which fall in the rotational frequency range
determined by us are: $\nu_1$ and $\nu_2$ in HD\,74560, $\nu_1$ and $\nu_4$ in HD\,123515,
$\nu_2$ in HD\,208057, $\nu_1$ and $\nu_2$ in HD\,28114, $\nu_1$ and $\nu_2$ in HD\,28475,
$\nu_1$ and $\nu_4$ in HD\,34798, all frequencies of  HD\,37151, $\nu_1$ and $\nu_2$ in HD\,45284,
all frequencies of HD\,143309, $\nu_1$ in HD\,179588, all frequencies of HD\,191295 and HD\,206540.
{We presume that most of these frequencies are of pulsational origin,
but it must be remembered that pulsational and rotational variability can coexist
in one star \citep[e.\,g.][]{Papics2012,Thoul2013} and even
spots alone could cause multi-periodic signal if the stellar surface rotates differentially \citep[e.\,g.][]{Degroote2011}.}
However, to decide  with certainty, a detailed abundance analysis
from high-resolution spectroscopy is necessary.

For some stars with spectroscopic data, viz. HD\,25558, HD\,74195 and HD\,182255,
we were unable to find solutions which reproduce the observed amplitude of the radial velocity variations
to within a 1$\sigma$ error. Therefore, we accepted solutions to within 2$\sigma$ or even
6$\sigma$ (see Table\,\ref{tab_results_all}).
This can be caused by model uncertainties (e.\,g.
model atmospheres, chemical composition, opacities, etc.
as well as the approach of computing stellar pulsations)
or observational data (e.\,g. underestimated errors of radial velocity amplitudes
or fundamental parameters constrained with too low precision).

Similarly, not always we were able to find a common range of the rotational velocity for all
modes of a given star. The stars in question are: HD\,123515, HD\,179588, HD\,208057, and for these stars
we followed the case of Model\#3 of our working example HD\,21071, i.e.,
for a given frequency we choose a mode with the range of $V_\mathrm{rot}$
closest to the range of $V_\mathrm{rot}$ obtained for other frequencies
observed in the star (they are printed in italic in Table\,\ref{tab_results_all}).

Finally, in some cases the identifications of the mode degree in
the zero-rotation approximation and the traditional approximation did not agree with each other.
An example is HD\,25558. For this star, both methods indicate that
$\nu_1$ is a dipole mode, but $\nu_3$ is a mode with $\ell$=1, 2 or 3
when the zero-rotation approximation is used or $\ell$=4, 6 when
the traditional approximation is used.
This is not surprising because rotation strongly modifies properties of pulsational modes of the SPB stars.
In particular, rotation changes the frequency domain of the unstable modes. This phenomenon is
especially pronounced for high values of $V_{\mathrm{rot}}$ and $|m|$.

In the last column of Table\,\ref{tab_results_all}, we listed for five stars
spectroscopic determinations of $\left( \ell\,,m\right)$ taken from the literature (De Cat et al. 2005).
In the case of HD\,24587, HD\,92287 and HD\,181558, our mode identification is in agreement
with the angular numbers found by De Cat et al. (2005) but for HD\,140873
and HD\,177863 the results are in contradiction. However,
if we look at solutions within 3$\sigma$ in $A_{V_{\rm rad}}$ for $\nu_1$ in HD\,177863, the modes (1,0)
and (1,+1) are also possible and the last one agrees with the determinations of De Cat et al (2005).
In the case of $\nu_1$ in HD\,140873, we are dealing with a similar situation but the agreement is achieved
within 4$\sigma$ in $A_{V_{\rm rad}}$ and the only additional solution is the mode (1,+1).
On the other hand, it should be noted that both stars are visual and spectroscopic
binaries and photometric as well as spectroscopic observations can be polluted by
the light from the companions, and different methods can be sensitive to this effect in
different ways. Moreover, if $\nu_1$ in HD\,177863
is tidally excited, then according to \citet{Willems_Aerts2002}
it should be a quadrupole sectoral mode.

Unfortunately, for HD\,21071, which is the best studied star by us in the sample, there are no
determinations of angular numbers in the literature from the spectroscopic
methods and we cannot compare our results. There is only the identification
of the harmonic degree from the photometric method. \citet{DeCat2007}
found that $\nu_1$ is a dipole mode what is in agreement with our result. For $\nu_2$, $\nu_3$ and $\nu_4$,
they found that these frequencies can be associated with dipole or quadrupole modes
and these results are also compatible with our identifications. The only discrepancy is that according to \citet{DeCat2007}
$\nu_2$ and $\nu_4$ can be also the $\ell=4$ mode.
In fact, this solution was found also by us but we rejected it because this mode was stable.

\section{Summary}
We obtained a mode identification or at least constraints on angular numbers ($\ell$, $m$)
for 31 SPB stars. Simultaneously, we derived bounds on the rotational
velocities of the stars we examined. 
It should be noted, that the results delivered by the presented method
depend on the model parameters. It would be very interesting to verify them on the SPB stars
for which the inclination angle
and/or rotation period are known from independent determinations.
We see such prospects in the forthcoming two-colour photometric data from the \emph{BRITE} satellites
and follow-up spectroscopic observations.

\begin{figure}
\centering
\includegraphics[angle=-90, width=\columnwidth]{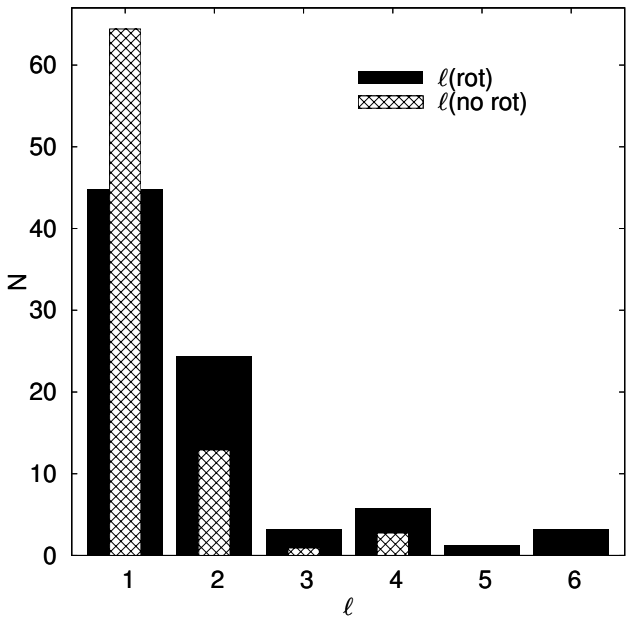}
\caption{The number of the modes with different values of the harmonic degree, $\ell$,
              obtained with (black filled bars) and without (hatched bars) the effects of rotation taken into account.\label{hist_l}}
\end{figure}

\begin{figure}
\centering
\includegraphics[angle=-90, width=\columnwidth]{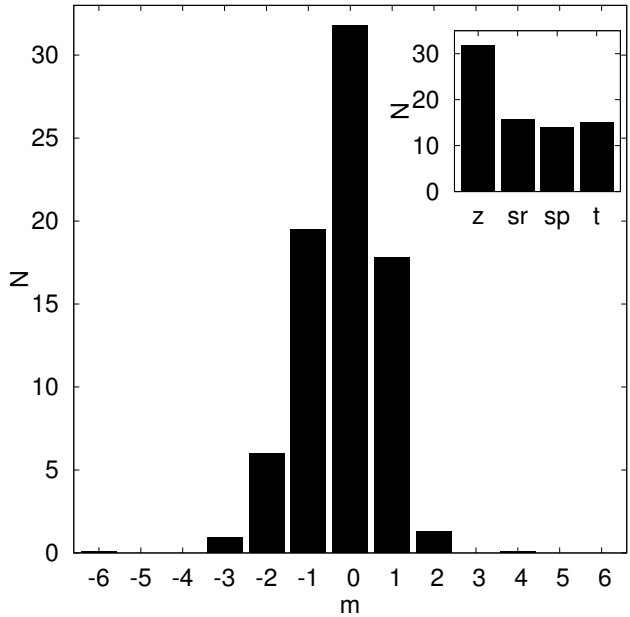}
\caption{The number of the modes with different values of the azimuthal order, $m$.
              The insert shows the histogram of the number of the zonal (z), sectoral retrograde (sr),
              sectoral prograde (sp) and tesseral (t) modes. \label{hist_m}}
\end{figure}

\begin{figure}
\centering
\includegraphics[angle=-90, width=\columnwidth]{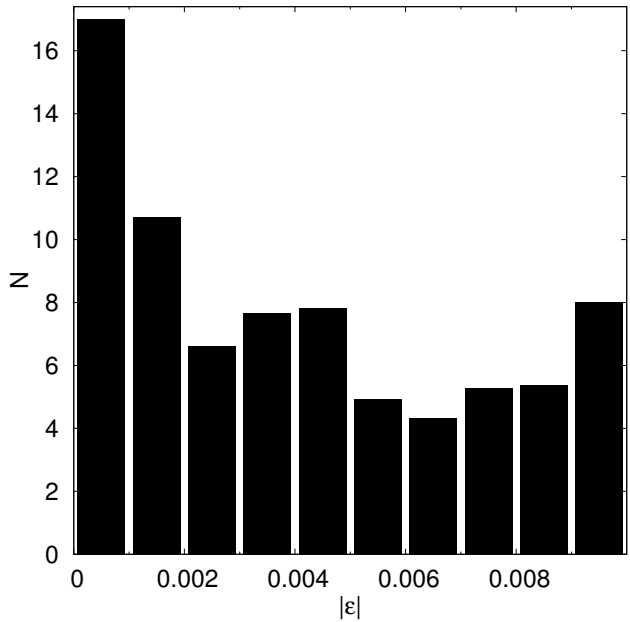}
\caption{Distribution of the values of $\varepsilon$ for the frequency peaks of the studied stars.\label{hist_eps}}
\end{figure}

\begin{figure}
\centering
\includegraphics[angle=-90, width=\columnwidth]{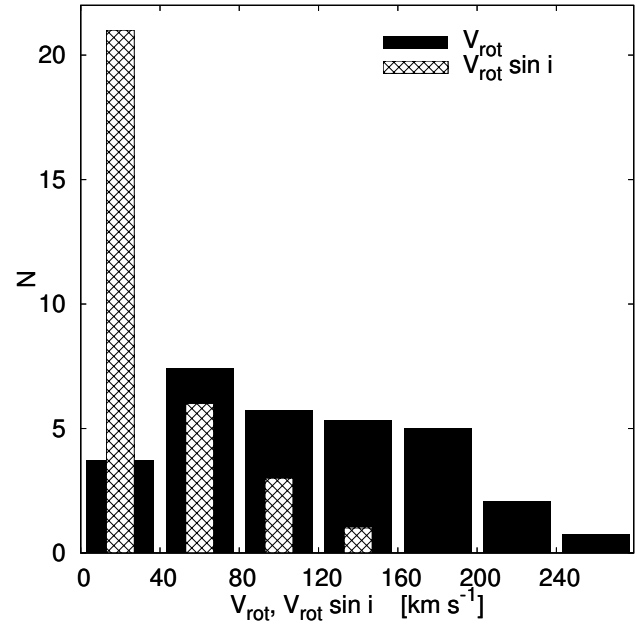}
\caption{Distribution of the values of ${V}_\mathrm{rot} \sin i$ (hatched bars) and the values of $V_{\rm rot}$ (black solid bars)
         we obtained simultaneously with the identification of $(\ell, ~m)$ for the studied stars.\label{hist_V}}
\end{figure}

\begin{figure}
\centering
\includegraphics[angle=-90, width=\columnwidth]{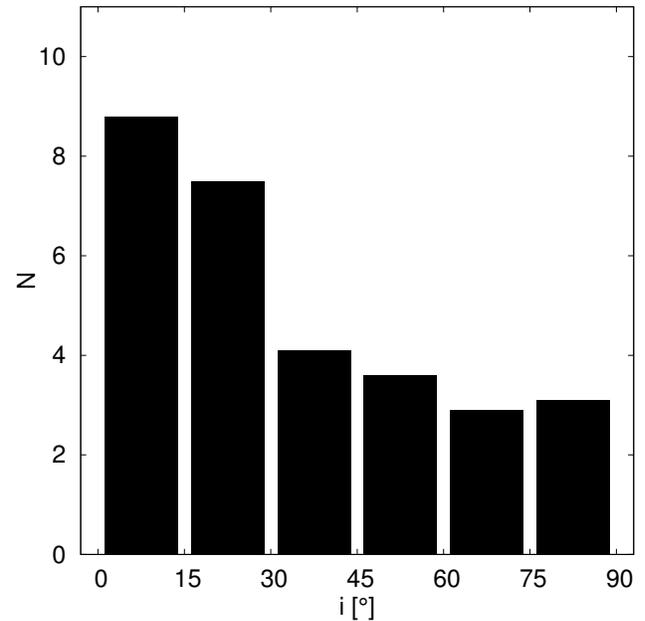}
\caption{Distribution of the inclination angle, $i$, for the stars of Table 1.\label{hist_i}}
\end{figure}

To summarize our results, we constructed histograms showing the distributions of the mode degree, $\ell$, (Fig.\,\ref{hist_l}),
azimuthal order, $m$ (Fig.\,\ref{hist_m}), the intrinsic mode amplitude, $\varepsilon$ (Fig.\,\ref{hist_eps}),
the values of  $V_\mathrm {{rot}}$ and $V_\mathrm {{rot}} \sin i$ (Fig.\,\ref{hist_V}),
and the inclination angle, $i$ (Fig.\,\ref{hist_i}).
Due to the fact that for some frequencies the identification was not unambiguous,
we shall now explain how we have counted the modes with a given value of the mode degree or
azimuthal order. In the case of the modes for which we found $j$ combinations of ($\ell$, $m$)
and $\ell$ or $m$ occurred $i$ times, we counted such angular numbers  $i/j$ times.
The histograms for $V_{\rm rot}$ and $i$ were constructed in a similar way.

When we neglect all effects of rotation, most of the modes
appeared to be dipole modes (see Fig.\,\ref{hist_l}). Some modes
were quadrupole modes and very few had $\ell$ equal to 3 or 4.
None of the frequencies had $\ell=5$ or 6.
If the effects of rotation were included, the number of the dipole modes decreased significantly
and number of the quadrupole modes was approximately doubled. We also obtained slightly larger
number of the modes with $\ell$ higher than 2.
Despite this, only very few modes had $\ell \ge 3$. One can see also that
there were fewer modes with  $\ell=3$ and 5 than modes with $\ell=4$ and 6.
This is because the odd degree modes have lower visibility
\citep[see][]{JDD2002, JDD2015}.

From Fig.\,\ref{hist_m} one can see that zonal modes are quite numerous amongst our sample of the 31 stars.
We have to admit that we do not have any explanation of that fact.
Maybe this is the case or it can result from model uncertainties and/or limitations of the method we applied.
The number of sectoral retrograde and prograde modes is about the same
and if we count them all, the number of the sectoral modes becomes only slightly smaller than
the number of the zonal modes (cf. the insert of Fig.\,\ref{hist_m}).
Some modes with higher values of $m$ also happen.

In the method we used, the values of the intrinsic mode amplitude, $\varepsilon$, are derived from observations (equation 3)
and used to constrain mode identification. The histogram for $\varepsilon$ is shown
in Fig.\,\ref{hist_eps}.
An interesting fact is that most values of $\varepsilon$ are much smaller than the adopted upper limit, $\varepsilon_{\rm max}=0.01$.
A similar result was obtained for the frequency peaks of $\mu$ Eri by \citet{JDD2015}.
In the case of our sample of stars, above 60 per cent of the frequency peaks have $\varepsilon$ below 0.005.
If we limit the sample to the stars with the radial velocity measurements, the number
of the frequency peaks with $\varepsilon<0.005$ increases to above 80 per cent.

In Fig.\,\ref{hist_V}, we present a histogram showing the distribution
of the values of the rotational velocity we obtained and
the projected rotational velocity. One can see that a substantial number of the programme  stars
have $V_\mathrm{{rot}}\sin i$ below 40 km s$^{-1}$.

One can observe a quick decrease of the number of stars with large values
of $V_\mathrm{{rot}}\sin i$, and there are no stars with $V_\mathrm{{rot}}\sin i$
higher than 160 km s$^{-1}$. The view is quite different when we consider the `true' value of the rotational velocity.
In the first place, stars rotating below 40 km s$^{-1}$ are not most numerous and
there are stars with rotational velocities from the whole range
of $V_\mathrm{{rot}}$. The number of stars with $V_\mathrm{{rot}}$
in the range 40--80 km s$^{-1}$ is greater by a factor of 2 than that of stars with the smallest
rotational velocities. Then the number of stars is
slowly decreasing with the increasing value of $V_\mathrm{{rot}}$.

{The distribution of the projected rotation velocities and rotation velocities of B-type stars
can be found in, e.\,g.  \citet{Huang2006}. Although in the cited paper it is shown that the
distribution of $V_\mathrm{{rot}} \sin i$ for the field and cluster stars differ slightly,
it is striking that the values of $V_\mathrm{{rot}} \sin i$ in our sample are
lower and the number of stars quickly decreases
with increasing $V_\mathrm{{rot}} \sin i$.
\citet{Huang2006} also showed that most numerous among B-type stars are those
rotating with the equatorial velocity equal to about 200 km s$^{-1}$. In our sample,
the most numerous are stars rotating in the range 40--80 km s$^{-1}$.
Such shift in distribution can suggest that SPB stars are indeed slower rotators than
the other B-type stars 
\citep[e.\,g.][]{Balona2010}.
As opposed to \citet{Huang2006}, \citet{Levato2013} found
that the distribution of $V_\mathrm{{rot}} \sin i$ as well as $V_\mathrm{{rot}}$ for
B-type stars is bimodal what is not seen in our sample.
However, we want to emphasize that our sample is too small to make any
certain conclusions.}

Finally, in Fig.\,\ref{hist_i} is presented a distribution of the inclination angle, $i$,
we derived for our sample of stars. One can see that about half of the stars are observed
at rather low inclination angle $i<30^\circ$.
{As shown by \citet{Abt2001}, the stellar rotational axes should be  distributed randomly
so one would expect rather the flat distribution of $i$ but again we have to remind
that we are dealing with a small sample and the shape of histogram can be accidental.
The other reason  could be} that we identified so many zonal modes which are best seen from the pole-on direction (cf. Fig.\,\ref{hist_m}).
The other explanation is that the assumption of uniform rotation is incorrect and the higher values of $V_{\rm rot}$
(implying the low values of $i$) correspond to the deeper layers which are better probed by high-order g modes.
Some support for this hypothesis is the fact that there is a comparable population of the sectoral modes
which are least cancelled when we look at the equator and there is no increase
in the number of stars with the inclination angle close to 90$^\circ$.
This is an interesting explanation but it demands
more detailed studies which go beyond the scope of the present paper.

The unambiguity of mode identification strongly depends on the quality of the photometric and/or spectroscopic data.
The quality of the data is especially important when the effects of rotation are included.
This is due to the fact that contrary to the
case of the zero-rotation approximation when we determined only one parameter, $\ell$,
in the case with rotation included
we have to obtain additionally the azimuthal order, $m$, and rotational velocity,
$V_\mathrm{rot}$.
The information contained in the amplitude of the radial velocity variations
helps to reduce significantly the number of possible solutions. Thus,
combined photometric and spectroscopic data are highly desirable.

In the case of HD\,123515, HD\,179588 and HD\,208057 as well as in Model\#3
{and Model\#5} of HD21071,
we failed finding a common range of the rotational velocity for all observed
frequencies. This can be explained in several ways.
First, the accuracy of our pulsational calculations may be insufficient
and the theory needs to be improved.
{Secondly, the formal errors of the amplitudes and phases of the light and radial velocity variations
were underestimated. Thirdly, the astrophysical parameters we used, $\log T_\mathrm{eff}$ and $\log L/\mathrm L_\odot$,
were not possible to be constrained with a high precision.
It is also possible that the adopted initial abundance of hydrogen, metallicity or overshooting
from the convective core made the starting models inappropriate.}
And again, it is possible that our assumption of rigid rotation is not satisfied,
so that different $V_\mathrm{rot}$ ranges for different frequency peaks
observed in a star are manifestations of differential rotation.
Perhaps all above-mentioned factors play a role simultaneously.

{Finally, close companions can disrupt photometric and
spectroscopic measurements and thus the mode identification.
Two of three stars with the high $\chi^2$ solutions,
HD\,53921 and HD\,179588, are visual binary or visual and spectroscopic binary, respectively.
A high value of $\chi^2$ can also suggest non-pulsational originate of variability.}

Nevertheless, we see prospects for seismic modelling of the SPB stars for which the best constraints
on the angular numbers, ($\ell,~m$) and the rotational velocity, $V_\mathrm{rot}$, were obtained.
Only in such cases one can try to attribute the radial order, $n$, to the observed frequency peaks.
In particular, HD\,21071 is a promising candidate. Our results of seismic modelling of this star
will be presented in a separate paper (Szewczuk \& Daszy\'nska-Daszkiewicz, in preparation).
{Preliminary results has been already presented in \citet{Szewczuk2015}.}

\section*{Acknowledgements}
We gratefully thank Mike Jerzykiewicz for his comments and carefully reading the manuscript.
{We are also indebted to an anonymous referee for her/his valuable remarks.}
WS was financially supported by the Polish NCN grant DEC-2012/05/N/ST9/03905 and JDD
by the Polish NCN grants 2011/01/M/ST9/05914,  2011/01/B/ST9/05448.
{Calculations have been carried out using resources provided by Wroc{\l}aw Centre
for Networking and Supercomputing (http://wcss.pl), grant No. 265}

\label{lastpage}

\end{document}